\documentclass{nature}
\usepackage{graphicx}
\usepackage{dcolumn}
\usepackage{bm}
\usepackage{amsmath}
\usepackage{amssymb}
\usepackage{color}
\usepackage{tikz}
\usepackage{caption}
\title{Experimentally Probing Topological Order and Its Breakdown via Modular Matrices}

\author{Zhihuang Luo$^{1,2}$, Jun Li$^{2}$, Zhaokai Li$^{1}$, Ling-Yan Hung${^{3,4,5}}^{*}$, Yidun Wan${^{4,5,6}}^{*}$, Xinhua Peng${^{1,7,8}}^{*}$,  \& Jiangfeng Du$^{1,7}$}

\begin{document}

\maketitle

\begin{affiliations}
 \item CAS Key Laboratory of Microscale Magnetic Resonance and Department of Modern Physics, University of Science and Technology of China, Hefei, Anhui 230026, China
 \item Beijing Computational Science Research Center, Beijing, 100094, China
%  \item Institute for Quantum Computing and Department of Physics and Astronomy, University of Waterloo, Waterloo N2L 3G1, Ontario, Canada
\item State Key Laboratory of Surface Physics and Department of Physics, Fudan University, 220
Handan Road, 200433 Shanghai, China
 \item Department of Physics and Center for Field Theory and Particle Physics, Fudan University,
220 Handan Road, 200433 Shanghai, China
\item Collaborative Innovation Center of Advanced Microstructures, Nanjing 210093, China
 \item Perimeter Institute for Theoretical Physics, Waterloo, Ontario N2L 2Y5, Canada
 \item Synergetic Innovation Center of Quantum Information and Quantum Physics, University of Science and Technology of China, Hefei, Anhui 230026, China
\item Synergetic Innovation Center for Quantum Effects and Applications, Hunan Normal University, Changsha 410081, China
\end{affiliations}

\begin{abstract}
  The modern conception of phases of matter has undergone  tremendous developments since the first observation of topologically ordered states in fractional quantum Hall systems in the 1980s. In this paper, we explore the question: How much detail of the physics of topological orders can in principle be observed using state of the art technologies? We find that using surprisingly little data, namely the toric code Hamiltonian in the presence of generic disorders and detuning from its exactly solvable point, the modular matrices -- characterizing anyonic statistics that are some of the most fundamental finger prints of topological orders -- can be reconstructed with very good accuracy solely by experimental means. This is a first experimental realization of these fundamental signatures of a topological order, a test of their robustness against perturbations, and a proof of principle -- that current technologies have attained the precision to identify phases of matter and, as such, probe an extended region of phase space around the soluble point before its breakdown. Given the special role of anyonic statistics in quantum computation, our work promises myriad applications both in probing and realistically harnessing these exotic phases of matter.
\end{abstract}

Landau's theory fails to describe many exotic phases of matter such as topological orders, where no symmetry breaking is involved \cite{Landau1937,Landau1950,Wen1990TO,Wenbook}. Topological orders are characterized by their  robust ground-state degeneracy and long-range entanglement \cite{Wen1990D,Chen2010E}. The fractional quantum Hall states are among the best known examples of such states \cite{Tsui1982}. It is well known that topological orders do not admit any local order parameters. Characterization of topological orders based on quantities such as topological entanglement entropy  \cite{TEE,TEE2} has also proved inadequate, as they often take the same value for different phases. It is thus a profound and paramount quest to single out the minimal set of topological observables that would uniquely identify a topological order.

\begin{figure}
  \centering
  \includegraphics[width=10cm]{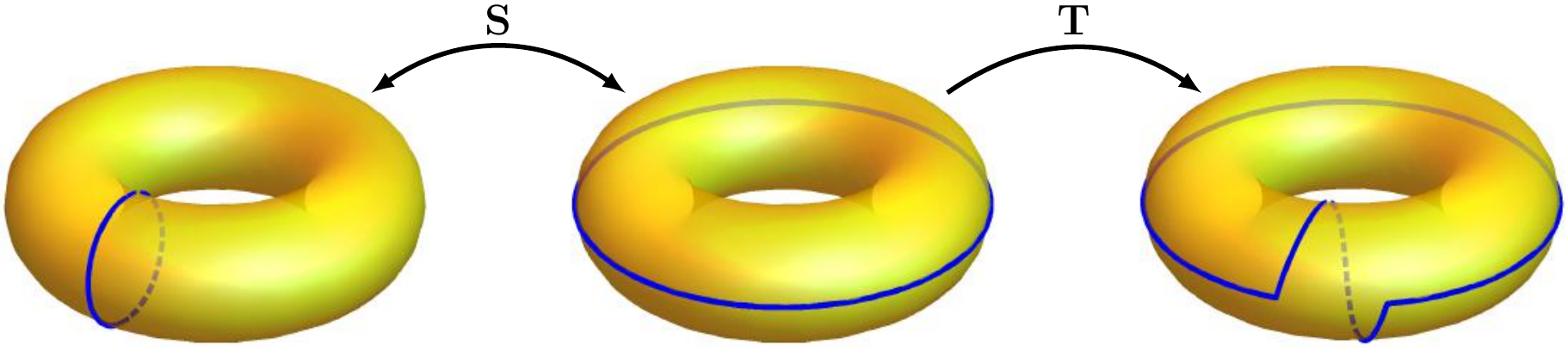}\\
  \caption{{\bf{Geometrical diagrams of the modular S and T transformations}}. They correspond to $\pi/2$ rotation and Dehn twist on a torus, respectively. }\label{fig_ST}
\end{figure}

Recent theoretical works have shown that modular matrices might help achieve precisely this goal \cite{Wen1993,WenST,Zhang2012,Bonderson2006,Vidal2013,Zaletel2013,WenPRB2015,WenPRL2015}. More specifically, imagine placing a two-dimensional system on a torus, which practically corresponds to periodic boundary conditions along two independent cycles. The $S$ and $T$ modular matrices record the Berry phases accumulated when one adiabatically deforms the system geometrically.  The $S$ matrix corresponds to a $\pi/2$ rotation,  and $T$ to a shear -- often called a Dehn twist -- of the fundamental region of the torus, as illustrated in Fig. \ref{fig_ST}.
Generically, a topological order has degenerate ground states on a torus, and the modular matrices furnish a non-Abelian representation of the modular group $SL(2,\mathbb{Z})$, which are thus instances of non-Abelian geometric phases. Like a fingerprint, modular matrices can be used to distinguish different topological orders, in fact uniquely for bosonic non-chiral phases. For example, the $\mathbb{Z}_2$ toric code order and the doubled semion order have different modular matrices\cite{Liu2014}, which also encode the anyonic statistics in these states. We review this in the Supplementary Information, along with some further details on the theoretical description of the modular matrices.

In principle, modular matrices can be obtained in a discrete model from ground-state wave-function overlap after appropriate transformations on the lattice. This has been employed as a numerical test to identify topological orders \cite{Zhang2012,Vidal2013,Liu2014,Jiang2014,WenPRB2015,WenPRL2015}. Another recent progress was also achieved by constructing string operators, which is highly non-trivial even numerically when the model is not exactly solvable \cite{Bridgeman2016}.
A concrete demonstration that these quantities can in fact be measured to sufficient accuracy using current experimental techniques would turn these pure theoretical discussions into realistic physical observables, opening up entirely new possibilities in experimental studies of these orders. In this paper, we are more ambitious than that: We study a family of spin Hamiltonians at different points in their phase space and measure the corresponding modular matrices. We show that without further theoretical input other than identifying  \emph{approximate} point-group symmetry of the Hamiltonian, the experiment proves itself capable of identifying the underlying topological order, in this case the $\mathbb{Z}_2$ topological order, by accurately recovering the modular matrices that stay exactly stationary before the analogue of phase transition-- more appropriately level crossing where higher levels cross the supposedly degenerate ground state subspace.\footnote{We clarify our meaning of a phase transition in detail in the Supplementary Information.}  We note that information about point-group symmetry is necessary because the modular transformations were implemented by permuting the lattice sites which would only preserve the degenerate ground state subspace if they were symmetries of the Hamiltonian. It is remarkable however that an approximate symmetry suffices, where the symmetry breaking scale is smaller than the topological mass gap.  The experiment can also probe finite regions of the phase diagram, and locate the maximal value
of the magnetic field where a transition occurs, beyond which the topological properties change abruptly, exhibiting in action the drastic phenomenology of topological orders that  has a Hall-plateau like flavour.

\begin{figure}
  \centering
  \includegraphics[width=7.5cm]{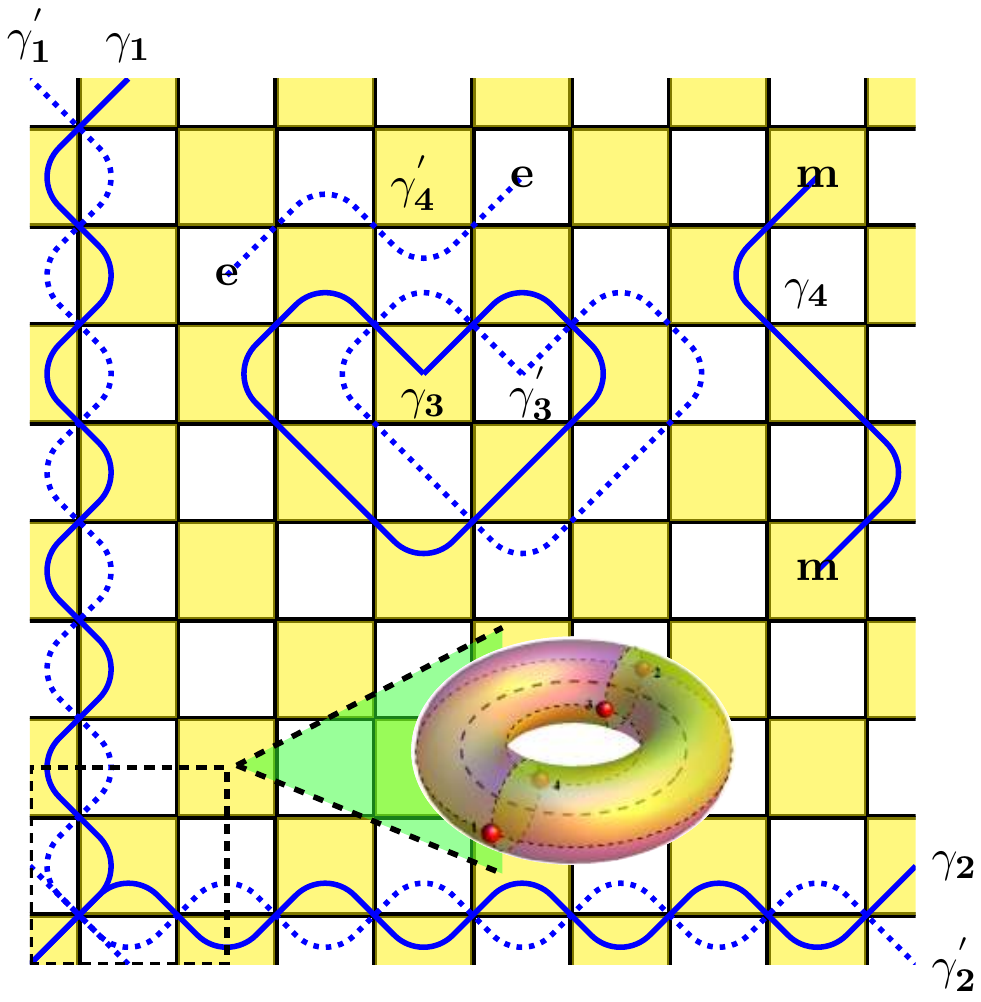}\\
  \caption{{\bf{Kitaev toric code model on a torus.}} When $N$ is even, there exist two sublattices denoted by white and yellow. The blue solid strings ($\gamma_1\sim\gamma_4$) and their dual dashed strings ($\gamma_1^{'}\sim\gamma_4^{'}$) are defined in the yellow sublattice and in the white sublattice, respectively. $e$ and $m$ represent the elementary excitations (anyons): electric charge and magnetic vortex, which are in pairs generated by open string operators. The black dashed box is a unit cell of square lattice and forms a torus under the periodic boundary condition. The red spheres on the torus represent spins.}\label{fig:model_1}
\end{figure}

%\section{Hamiltonian} \label{sec:Hamiltonian}
Let us first describe target Hamiltonian studied in this experiment, which reads
\begin{equation} \label{systemH}
\hat{H}_{\text{T}} = \hat{H}_{\mathbb{Z}_2} - h \sum_i \hat{\sigma}_i^z -  \sum_i{} \epsilon_i\,\, \hat{\sigma}_i^z,
\end{equation}
where on each site $i$ of the $N\times N$ square lattice resides a spin $1/2$, and $\hat{\sigma}^{(x,y,z)}_i$ are the Pauli matrices acting on site $i$. Let us explain the special role of each term in equation \eqref{systemH} separately. The first term is the Hamiltonian of the Kitaev $\mathbb{Z}_2$ toric code model \cite{WenPRL2003}, which in its simplest version reads
\begin{equation}\label{Ham}
    \hat{H}_{\mathbb{Z}_2}=\sum_{\text{white plaquettes}}X_p+\sum_{\text{yellow plaquettes}}Z_{p}.
\end{equation}
Here $X_p=\prod_{j\in\partial p}\hat{\sigma}_j^x, Z_p=\prod_{j\in\partial p}\hat{\sigma}_j^z$ are the plaquette operators that act on four spins surrounding a plaquette of $p$. This is illustrated in Fig. \ref{fig:model_1}.

It is well known that $H_{\mathbb{Z}_2}$ is exactly solvable because  $[X_p,Z_{p'}]=0$ for all $p$ and $p'$. The ground state subspace is given by
\begin{equation}\label{}
  \mathcal{L}=\{|\psi_g\rangle\in\mathcal{H}:X_p|\psi_g\rangle=Z_p|\psi_g\rangle=-|\psi_g\rangle \text{ for all } p\}.
\end{equation}
The ground-state degeneracy is $D=2^{2\textsf{g}}$, where $\textsf{g}$ is the genus number of the Riemann surface where the system resides. If the Riemann surface has genus $\textsf{g}$, we can define $2\textsf{g}$ non-contractible loops that connect to different ground states in $\mathcal{L}$. Particularly, on a torus the model has four-fold degenerate ground states. The Kitaev model  is the infrared fixed point model that describes the $\mathbb{Z}_2$ toric code order. The $\mathbb{Z}_2$ topological order however exists in a finite region of the phase space of the model. Particularly, a homogenous magnetic field, corresponding to the second term in Eq. (\ref{systemH}), can be added while preserving the order. Sufficiently large magnetic field would cause the system undergo a transition to a trivial phase. This has been studied quite extensively in the literature\cite{Nayak2007,Vidal2011}. The modular matrices are indeed invariant throughout the entire phase space before the transition occurs.  To showcase the robustness of the order stemming from its topological nature, we would like to break all translation symmetry in the system, by introducing a random magnetic field at each site, which yields the third term of Eq. (\ref{systemH}).

In the experiment, we consider a unit cell (i.e., a $2\times 2$ square lattice) of the torus as our test system. The ground states of the Kitaev toric code model are fixed point wave-functions, whose topological properties are independent of lattice size\cite{WenPRL2003}. Under the periodic boundary condition, the total Hamiltonian reduces to
\begin{equation}\label{Ham4}
\hat{H}_{\text{T}}^4=2(\hat{\sigma}_1^x\hat{\sigma}_2^x\hat{\sigma}_3^x\hat{\sigma}_4^x+\hat{\sigma}_1^z\hat{\sigma}_2^z\hat{\sigma}_3^z\hat{\sigma}_4^z) - h\sum_{i=1}^4 \hat{\sigma}_i^z-\sum_{i=1}^4 \epsilon_i \hat{\sigma}^z_i.
\end{equation}
The energy levels of the system as a function of $h$ is plotted in Fig. \ref{fig:5}a. The level crossing point is located at $h_c=2\sqrt{10}/3$. The disorders are randomly generated, with a strength  $\epsilon_i$ in the range $[-0.05,0.05]$. There is a significant hierarchy in energy gaps, between the splittings among the four lowest states in the ground state subspace and the ``topological'' gap. This feature is supposedly more pronounced in the thermodynamic limit. Therefore, the above definition of transitions for such small system asymptotes to the true phase transition in the thermodynamic limit. Let us emphasize here again that phase transitions for a small system is not an entirely well-defined concept. The best analogue of a phase transition in a small system is level crossing, which the experiment is essentially detecting. A thorough discussion of this delicate issue is relegated to the supplementary information. We are going to make measurements over four different choices of $h $ in the range $0\leq h<2.5$.

\begin{figure}
  \centering
  \includegraphics[width=12cm]{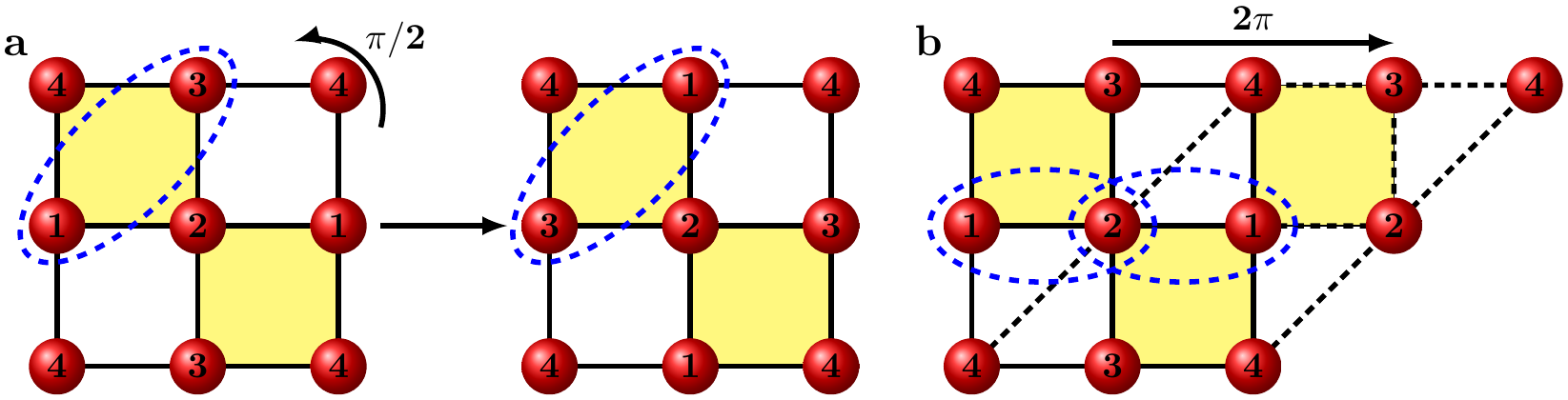}\\
  \caption{{\bf{Physical realizations of the modular S and T transformations on a $\mathbf{2\times 2}$ torus.}} {\bf{a}}, S: $\pi/2$ rotation equivalent to a $\text{SWAP}_{13}$ operation, and {\bf{b}}, T: a shear equivalent to a $\text{SWAP}_{12}$ operation.}\label{fig:st}
\end{figure}

Figures \ref{fig:st}a and \ref{fig:st}b clearly illustrate that the $\pi/2$ rotation and Dehn twist are equivalent to the experimental operations of $\text{SWAP}_{13}$ and $\text{SWAP}_{12}$, respectively, which keep the toric code Hamiltonian in a homogenous magnetic field invariant. Here $\text{SWAP}_{ij}$ is the operation of swapping spin $i$ and spin $j$. It is also necessary to emphasize that although we only consider a unit cell as our testing system, the proposal for measuring $S$ and $T$ matrices is scalable to larger systems. The details on experimentally implementing $S$ and $T$ transformations and their complexity analysis are given in the Supplementary Information. From the analysis, one can see that the measurement proposal acquires only polynomial complexity and is thus efficient.

Now we turn to the experimental realization to directly measure the modular matrices of the $\mathbb{Z}_2$ topological order. With the well-developed control technology \cite{Ray2008}, nuclear magnetic resonance (NMR) has been widely utilized for many of the first demonstrations in quantum simulation \cite{Feynman1982}. In the experiment we employed three $^{19}$F spins and two $^{1}$H spins of 1-bromo-2,4,5-trifluorobenzene partially oriented in liquid crystal N-(4-methoxybenzaldehyde)-4-butylaniline (MBBA) as a 5-qubit NMR simulator \cite{Mahesh2014}. The molecular structure and the labeled qubits are shown in Fig. \ref{fig:sample}a. The first $^{19}$F spin (labeled by 0) is used as a probe qubit, and the rest of the spins (labeled by $1\sim4$) constitute the  4-qubit  quantum register to simulate the system of a $2\times 2$ spin lattice. The experiment was carried out at 303 K on a Bruker AV-400 spectrometer. In our molecule, the effective couplings between nuclear spins originate from partially averaged dipolar interactions $D_{jk}$ (DD-couplings)  and scalar interactions $J_{jk}$ (J-couplings). Since the chemical shift difference in each pair of spins is much higher than the effective coupling strength, the $x$ and $y$ components in DD-coupling interaction can be ignored by secular approximation \cite{Dong1997}. Therefore, the effective Hamiltonian of this 5-qubit system in a doubly rotating frame is
\begin{equation}\label{}
    \hat{H}_{\text{NMR}}=\sum_{j=0}^4\pi\nu_j\hat{\sigma}_j^z+\sum_{ j<k, =0}^4\frac{\pi}{2}(J_{jk}+2D_{jk})\hat{\sigma}_j^z\hat{\sigma}_k^z,
\end{equation}
with the related parameters shown in Fig. \ref{fig:sample}b.

It is one of the main purposes in developing quantum simulators to obtain and subsequently measure the ground states of some given Hamiltonian \emph{dynamically}, thereby solving otherwise (classically) computationally challenging problem that often do not have analytic solutions. To ask the simulator to solve our problem at hand -- obtaining the degenerate ground states of the given Hamiltonian without prior analytical input, we come up with the method -- \emph{random adiabatic evolution}.  To put it simply, we randomly generate simple Hamiltonians as starting point and we adiabatically evolve the system to the target Hamiltonian. This would generically prepare dynamically all the linearly independent ground states with only a few trials.  This can be contrasted with prior numerical or experimental work that make use of the string operators \cite{Bridgeman2016,Luo2016}. The current strategy highlights the strengths of the NMR system, substituting as much analytical insight as possible by experimental maneuver.

The quantum circuit for randomly preparing linear-independent ground states is shown in Fig. \ref{fig:sample}c. Using line-selective approach \cite{Peng2001}, the quantum system was first prepared in the initial pseudo-pure state (PPS):
$\hat{\rho}_{00000}=\frac{1-\epsilon}{32}\bold{I}+\epsilon |00000\rangle\langle00000|$, with $\bold{I}$ representing $32\times32$ identity operator and $\epsilon\approx 10^{-5}$ the polarization. Here we introduced a probe qubit for the interferometry, which is initialized in the superposition state (i.e., $\frac{1}{\sqrt{2}}(|0\rangle_0+|1\rangle_0)$) by a Hadamard gate. In Fig. \ref{fig:sample}c, $\text{APGSi}: |0000\rangle_{1234}\mapsto|\psi_i^{\text{rd}}\rangle$ and $\text{APGSj}: |0000\rangle_{1234}\mapsto|\psi_j^{\text{rd}}\rangle$, where APGS stands for \emph{random adiabatic evolution}. Consider the following time-dependent Hamiltonian:
\begin{equation}\label{Had}
    \hat{H}(t)=[1-s(t)]\hat{H}_{\text{rd}}+s(t)\hat{H}_{\text{T}}^4,
\end{equation}
where $\hat{H}_{\text{rd}} = \sum_{i}^4\sum_{\alpha\in x,y,z} C_i^{\alpha} \hat{\sigma}^{\alpha}_i$ and $C^{\alpha}_i$ are randomly generated coefficients between $[-1,1]$.
To obtain the ground state of our target Hamiltonian $\hat{H}_{\text{T}}^4$, we first prepare the ground state of a simpler local Hamiltonian $\hat{H}_{\text{rd}}$, which can be readily implemented only using single-qubit rotations. The system is then evolved adiabatically by varying the parameter function $s(t)$ slowly enough from $0$ at $t=0$ to $1$ at $t=T$, where $s(t)$ was interpolated linearly with $M=100$ discretized steps, and the duration of each step is $\tau=T/M$. According to the adiabatic theorem \cite{Messiah1976}, the total evolution time obeys $T=O(1/\Delta E_{\text{min}})$, where $\Delta E_{\text{min}}$ is the minimum energy gap between four almost degenerate ground states and the first excited state encountered along the adiabatic evolution. The rate of change of $s(t)$ is so chosen that there is certain probability of excitation in the entire ground-state subspace.  Let us pause here and comment that the possibility of adiabatic evolution as a means to ground state preparation in topological orders were first proposed and studied in \cite{Hamma}, particularly for the detuned toric code model as in the current paper.  One crucial observation is that even in the thermodynamic limit and at couplings close to the topological phase transition, the adiabatic time scale $T$ needs only to scale at worst as $\sqrt{n}$ where $n$ is the total number of spins in the system to avoid jumping between the ground state subspace which is shown to be optimal. Moreover error following from excitations above the topological gap $\Delta$ can be made to be smaller than arbitrary polynomial  $(T\Delta)^N$ . Starting from different random Hamiltonian $\hat{H}_{\text{rd}}$, we can search out four linearly independent ground states by measurement results on the probe qubit, i.e., $\text{Tr}(\hat{\rho}_f\hat{\sigma}^{+}_0)=\frac{1}{2}\langle\psi_i^{\text{rd}}|\psi_j^{\text{rd}}\rangle$, where $\hat{\sigma}^{+}_0=|0\rangle_0\langle1|$ is the observable operator of the probe qubit (see method in detail). Let us remark that we can sensibly select four states out of the entire spectrum despite generically unavoidable finite size splittings among the ground state subspace is not an accident. There is a hierarchy between the finite size splitting and the topological gap controlled by the couplings, as shown in Fig. \ref{fig:5}a. This feature would only become more pronounced as the system size gets bigger, which is demonstrated in \cite{Nayak2007}. The experimental details can be seen in Supplementary Information.

Therefore, all elements of the modular matrices in four randomly generated linearly independent ground states (i.e.,  $\langle\psi_i^{\text{rd}}|S|\psi_j^{\text{rd}}\rangle$ or $\langle\psi_i^{\text{rd}}|T|\psi_j^{\text{rd}}\rangle$) can be subsequently obtained after the quantum circuit in Fig. \ref{fig:sample}d, directly from measuring the probe qubit. Note that the $S$ and $T$ matrices obtained in the random basis are not in their standard forms. In the standard form (labeled by $\langle\phi_i^{\text{std}}|S|\phi_j^{\text{std}}\rangle$ and $\langle\phi_i^{\text{std}}|T|\phi_j^{\text{std}}\rangle$), the basis states should correspond to anyons that diagonalize the $T$ matrix. To recover the standard form, we first construct the matrices in four orthogonal basis (denoted as $\langle\phi_i|S|\phi_j\rangle$ and $\langle\phi_i|T|\phi_j\rangle$) from the $S$, $T$ matrices in the random basis. Then in principle one can follow an algorithm proposed in Ref.\cite{Liu2014} to recover the standard basis. The recovering procedure can be summarized as follows: $\langle\psi_i^{\text{rd}}|S|\psi_j^{\text{rd}}\rangle\rightarrow\langle\phi_i|S|\phi_j\rangle\rightarrow\langle\phi_i^{\text{std}}|S|\phi_j^{\text{std}}\rangle$ and $\langle\psi_i^{\text{rd}}|T|\psi_j^{\text{rd}}\rangle\rightarrow\langle\phi_i|T|\phi_j\rangle\rightarrow\langle\phi_i^{\text{std}}|T|\phi_j^{\text{std}}\rangle$ (see Section Method below for detail). The main experimental results of standard $S$, $T$ matrices in four different magnetic fields are shown in Fig. \ref{fig:5}b. The values stay practically constant throughout $0\leq h <h_c$, which match to very high accuracy with those of the ``$\mathbb{Z}_2$ toric code modular tensor category'' \cite{Rowell2009}, in other words the $\mathbb{Z}_2$ toric code order. Immediately beyond the transition point, the optimization procedure returns a pair of modular matrices that differ significantly from the theoretical values of  the $\mathbb{Z}_2$ order. Importantly, they cease to be unitary because the invariant ground state subspace has changed across the transition and an arbitrary projection to the lowest four states do not preserve unitarity.
Our experiment therefore serves as a first experiment to identify topological order and detect its breakdown, which also give a clear and first experimental evidence that the topological phase, or at least its topological signatures of the toric code exists in an extended region of phase space around the soluble point.

The $^{19}$F spectra for measuring all elements of modular $S$ and $T$ matrices in the random basis are listed in Supplementary Information. Each element of the modular matrices was obtained directly from the integral intensity of all peaks in each spectrum. The final signal attenuated a lot due to the decoherence effect, which should be normalized. To estimate the decoherence, we considered the generalized amplitude damping and the phase damping acting on different spins \cite{Chuang2001}. The overall attenuation factor derived from the numerical simulation of the dynamical process was $\eta=0.2376$ and applied to normalize the final signals. The experimental errors mainly originated from the imperfection of the shaped pulses (about $1\%$) and the statistical fluctuation of signal strength (around $2\%$).

Although our experiments are performed on a four-qubit system, the scheme proposed here to measure the modular S and T matrices of a topological order and detect its existing range is in principle applicable to larger and more generic systems. As discussed above, the adiabatic method is shown to be optimal, with an adiabatic time scale $T$ that scales only at worst as $\sqrt{n}$ for a system of $n$ spins \cite{Hamma}. Another feature of topological orders that work to our advantage is its robustness that is made even more pronounced for larger system size. As system size grows, the energy splitting within the ground state subspace decreases exponentially with system size, while the topological gap remains constant in the limit. Such robustness is underlined in \cite{Nayak2007}, where it is demonstrated that the energy splitting among the ground state subspace diminishes quickly as system size increases, and that thermal excitation above the topological gap remains exponentially suppressed all the way up to phase transition as the detuning parameter $h$ is increased in the $\mathbb{Z}_2$ topological order. Therefore the choice of $T$ needs not be fine-tuned using detailed knowledge of the energy spectrum which is unavailable in general. Rather, we only need a rough estimate of the size of the topological gap which can be estimated from the values of the coupling. In this respect, finite size splitting effect of the topological ground state subspace may in fact be put to good use-- by picking a sufficiently large $T$ so that $1/T$ is smaller than all the energy gaps in the system, we will find the adiabatic method outputting to us the same state independently of the random starting Hamiltonian -- simulations beyond level crossing where the energy splitting gets significantly larger indeed confirms such an expectation. In other words, the random method itself is capable of discovering for us hierarchies in the energy gaps, and allowing us to make the case for selecting the lowest four states as part of the ground state subspace despite unavoidable level splitting in a finite size system. At the same time,  complexity of the quantum gates implementing S and T transformation only scale  polynomially with system size, and that our numerical method recovering the standard basis has no dependence on system size, all works to our advantage in the large system size limit. These are supporting evidence that future applications in larger systems is promising.  We will relegate more detailed comparison with larger system size in the supplementary information.

In summary, we presented the first proof-in-principle experimental identification of $\mathbb{Z}_2$ topological order by measuring the modular $S$ and $T$ matrices for a given Hamiltonian with minimal analytical input from state preparation to measurements. The only analytical input is in the approximate lattice symmetry of the system. This is a major improvement that puts the experiment to full use in finding its own ground state thereby solving the model, without requiring string operators that are not easily acquired for a generic model not exactly solvable.
 By recovering the modular matrices, some interesting properties such as quasiparticle statistics can be obtained from the resulting matrices. The measurement is shown to be robust against small local disorders in the form of an inhomogeneous magnetic field that breaks all accidental symmetries. The method also allows us to explore regions of the phase diagram in which a finite homogeneous magnetic field varies, and locate the phase transition, across which the modular matrices jump discontinuously.  Given the utility of the modular matrices in uniquely characterizing at least non-chiral bosonic 2+1 dimensional topological orders, the success of our NMR systems opens up future experimental avenues towards identifying topological orders whose Hamiltonians may not be exactly solvable. Our method is suitable not only for NMR systems, but also will work well in physical systems for quantum computers, such as superconductors and trapped ions. It will be interesting to generalize our measurement method to characterize other topological phases and their phase transitions.

\section*{Methods}

\subsection{Randomly preparing linearly independent ground states}

Starting from different initial Hamiltonians $\hat{H}_{\text{rd}}$ in \emph{random adiabatic evolution}, we can sequentially prepare a series of ground states denoted as $\{|\psi_1^{\text{rd}}\rangle,|\psi_2^{\text{rd}}\rangle,\cdots,|\psi_n^{\text{rd}}\rangle\}$, which can always be rewritten in the following form:
\begin{eqnarray}\label{eq:psi2phi}
  |\psi_1^{\text{rd}}\rangle &=& |\phi_1\rangle, \\ \nonumber
  |\psi_2^{\text{rd}}\rangle &=& a_2^1|\phi_1\rangle+a_2^2|\phi_2\rangle, \\ \nonumber
  \cdots & & \cdots \\ \nonumber
  |\psi_n^{\text{rd}}\rangle &=& a_n^1|\phi_1\rangle +a_n^2|\phi_2\rangle+a_n^3|\phi_3\rangle+\cdots+a_n^n|\phi_n\rangle,
\end{eqnarray}
where $\{|\phi_1\rangle,|\phi_2\rangle,\cdots,|\phi_n\rangle\}$ is an orthogonal basis and all $a_k^k$s are defined in the real number interval by removing the phase into $|\phi_k\rangle$. Using the quantum circuit in Fig. \ref{fig:sample}c, we measure the overlaps of these randomly generated ground states on the probe qubit, namely, $\langle\psi_i^{\text{rd}}|\psi_j^{\text{rd}}\rangle, i<j\in [2,n]$. The coefficients $a_i^j$ can be easily solved only from these overlaps.
If
\begin{equation}  \label{eq:error_tolerance}
a_k^k\gg|\xi_k|
\end{equation}
$|\psi_k^{\text{rd}}\rangle$ is linearly independent from $|\psi_{k'}^{\text{rd}}\rangle$ for $k'<k$; Otherwise, repeat the \emph{random adiabatic evolution} until finding all linearly independent ground states.   Here $\xi_k$ admits two contributions:
$\xi_k = |\chi| + k |\epsilon|. $
 First it is the readout error in the experiment which is $\chi$, and $\epsilon$, which is the overlap of the prepared states with the excited states beyond the supposed ground state subspace. The presence of non-vanishing $\epsilon$ means that
\begin{equation} \label{eq:pollute}
|\psi^{\textrm{rd}}_k\rangle =  \sum_{i} b_i |\tilde\psi_{g_i} \rangle +  \epsilon |\psi_{\textrm{ex}}\rangle,
\end{equation}
where $|\tilde\psi_{g_i}\rangle$ corresponds to states in the ground state subspace for the detuned model, and $|\psi_{\textrm{ex} }\rangle$ are states beyond the ground state subspace. Following the procedure using (\ref{eq:psi2phi}), it would suggest that $|\phi_k\rangle$ also takes a form like (\ref{eq:pollute}), provided that (\ref{eq:error_tolerance}) is satisfied. i.e. $| \phi_k \rangle$ remains mostly within the ground state subspace, and the contribution of the excited states remain linear in $\epsilon$. Far away from the transition point $h_c$, $\epsilon \le 10^{-3}$, and so $\phi_k$ remains safely within the ground state subspace. As $h_c$ is approached, $\epsilon$ can get  $>$0.2. As a result equation (\ref{eq:error_tolerance}) cannot in general be satisfied, and a drastic change in the modular matrix would be observed. A proper adiabatic time scale $T$ should thus be chosen such that $\epsilon$ remains sufficiently small as $h$ approaches $h_c$. An optimal $T$ can be chosen when the finite size splitting among the ground states is much smaller than the topological gap.

There are four linearly independent ground states in our experiment, which can usually be achieved within random 100 times. The experimental overlaps and coefficients are included in the Supplementary Information.

\subsection{Recovering standard modular matrices}

\begin{figure}
  \centering
  \includegraphics[width=13cm]{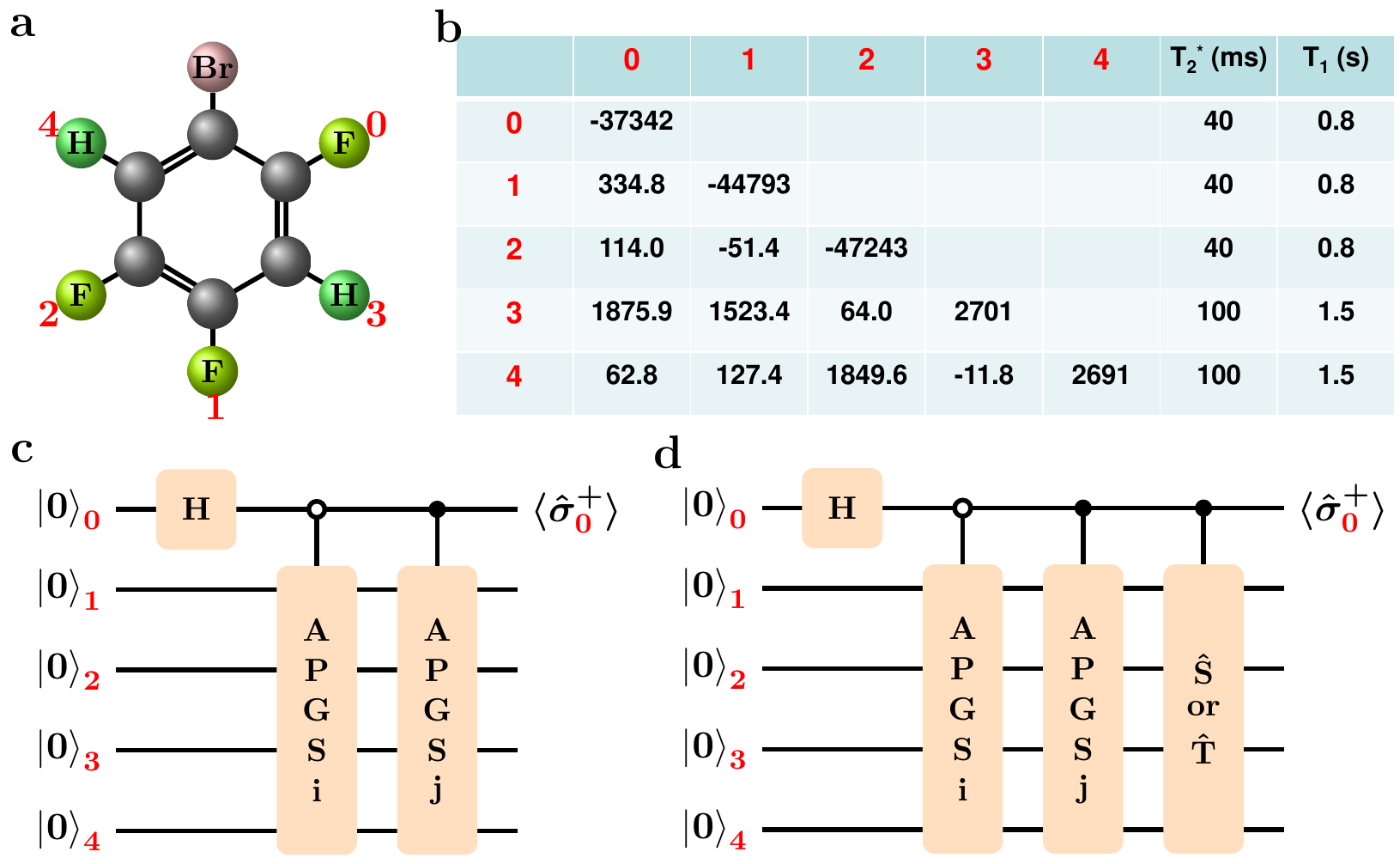}\\
  \caption{{\bf{Physical system and quantum circuit for measuring the modular matrices.}} {\bf{a}}, Molecular structure of the 1-bromo-2,4,5-trifluorobenzene. {\bf{b}}, Relevant parameters measured. The diagonal and off-diagonal elements represent the chemical shifts $\nu_i$ and effective coupling constants ($J_{jk}+2D_{jk}$) in units of Hz, respectively. {\bf{c}}, Quantum circuit for randomly preparing linear-independent ground states. H stands for the Hadamard gate acting on the probe qubit. The system (spins $1\sim 4$) is adiabatically prepared into a random ground state $|\psi_j^{\text{rd}}\rangle$, labeled by APGSj. {\bf{d}}, Quantum circuit for measuring the modular S and T matrices. Each element of S or T can be extracted by measuring the probe qubit.}\label{fig:sample}
\end{figure}

\begin{figure}
  \centering
  \includegraphics[width=13cm]{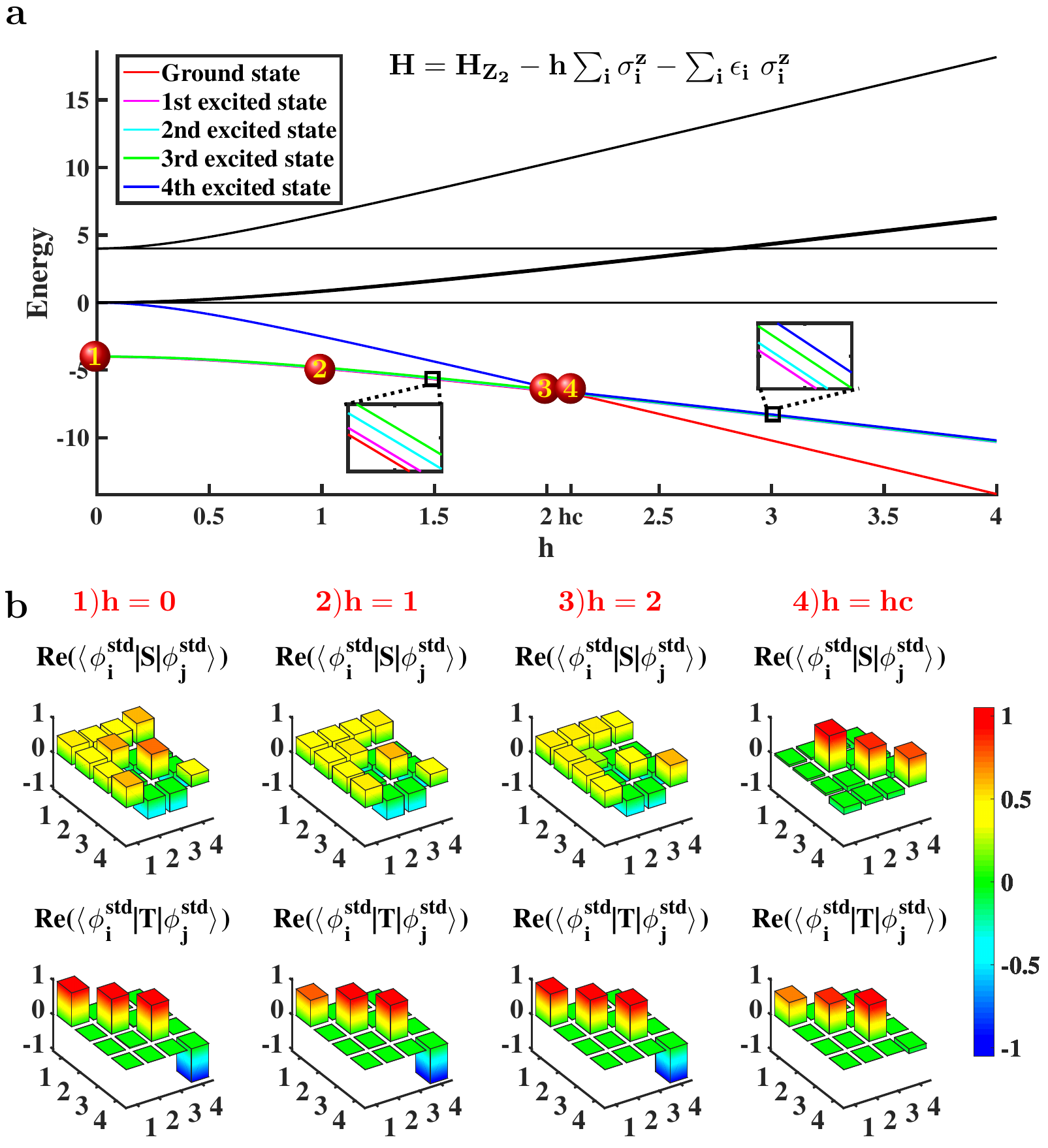}\\
  \caption{{\bf{Energy-level diagram of Hamiltonian (\ref{Ham4}) and experimental results of standard modular S and T matrices in different homogenous magnetic field.}} In {\bf{a,}} level crossing point is located at $h_c=2\sqrt{10}/3$. The small disorder $\epsilon_i$-term breaks the translation symmetry of Hamiltonian and open the four-fold degenerate ground states slightly. In {\bf{b,}} the resulting matrices stay the same values with high accuracy before the level crossing point and begin to be different drastically after that point. Beyond the ``phase transition'', the energy gaps quickly get much larger than our adiabatic time scale. With the loss of energy hierarchies, the random adiabatic method returns the same ground state every time.}\label{fig:5}
\end{figure}

According to equation (\ref{eq:psi2phi}), we have
\begin{equation}\label{}
    \langle\psi_i^{\text{rd}}|\mathcal{O}|\psi_j^{\text{rd}}\rangle=\sum_k A_{ik}\langle\phi_k|\mathcal{O}|\phi_j\rangle,
\end{equation}
where $\mathcal{O}= S$ or $T$, and $A_{ik}$ is the transformation matrix that have been obtained by the procedure above. Once all elements $\langle\psi_i^{\text{rd}}|\mathcal{O}|\psi_j^{\text{rd}}\rangle$ are measured directly from Fig. \ref{fig:sample}d, we can construct $\langle\phi_i|\mathcal{O}|\phi_j\rangle$ for all $i,j$. The results of  the $S$ and $T$ matrices in the random basis and in the orthogonal basis are shown in Supplementary Information. Note that the experimental errors inevitably lead to $S$ and $T$ matrices non-normal. To diagonalize the $T$ matrix for recovering the standard form in the next step, it is necessary to constrain the matrices in the orthogonal basis into normal matrices, which can be realized by optimal searching algorithm, i.e., to find the normal matrices that is the closest to experimental matrices. Actually, the normal constraint condition also makes the experimental data more physical. Then in principle one can follow an algorithm proposed in Ref. \cite{Liu2014} to recover the standard basis. Here we simplified the procedure slightly given the specific problem at hand. First, like in  Ref.\cite{Liu2014} we first diagonalize the $T$ matrix.  As we will see in the detailed discussion of the experimental data, the diagonalization of $T$, whose eigenvalues are the self-statistics of the anyons in a topological order, immediately suggests that the phase carries three bosonic anyons and one fermion. One can isolate the fermionic basis corresponding to the eigenvector of $T$ with -1 eigenvalue and recover the correct basis for the bosonic anyons by requiring the following: i)There exists a boson that corresponds to the trivial sector and which has trivial statistics with all other anyons; ii)They are orthogonal. A numerical optimization quickly converges, recovering the pair of $S$ and $T$ matrices in their standard form of the $\mathbb{Z}_2$ topological phase before level crossing occurs. Across the phase transition, one readily sees a jump in the modular matrices, which also cease to be unitary. This is because in the absence of an invariant ground state subspace under the $S$ and $T$ transformations, which is the case of a trivial phase or a different order where a different subspace is preserved, an arbitrary projection to some subspace does not preserve unitarity. This allows a precise identification of the phase transition.

%% Here is the endmatter stuff: Supplementary Info, etc.
%% Use \item's to separate, default label is "Acknowledgements"

\begin{addendum}
 \item We thank Xiaogang Wen and Bei Zeng for helpful discussions. This work is supported by NKBRP (2013CB921800 and 2014CB848700), the National Science Fund for Distinguished Young Scholars (11425523), the National Natural Science Foundation of China (Grants No. 11375167, No. 11227901, No. 11575173, and No. 91021005), the Strategic Priority Research Program (B) of the CAS (Grant No. XDB01030400), and RFDP (20113402110044). YW acknowledges the support from the John Templeton foundation No. 39901. This research was supported in part by Perimeter Institute for Theoretical Physics. Research at Perimeter Institute is supported by the Government of Canada through the Department of Innovation, Science and Economic Development Canada and by the Province of Ontario through the Ministry of Research, Innovation and Science.  LYH would like to acknowledge support by the Thousand Young Talents Program, and Fudan University.
 \item[Author Contributions] X. P. initiated the project. L. H. and Y. W. formulated the theory. X. P. and Z. L. designed the experiment. Z. L. and L. H. performed the calculation. Z. L. carried out the experiment and analyzed the data. X. P. and J. D. supervised the experiment. Z. L and L. H. wrote the draft. All authors contributed to discussing the results and writing the manuscript.
 \item[Competing Interests] The authors declare that they have no competing financial interests.
 \item[Correspondence] Correspondence and requests for materials
should be addressed to X. H. P.(xhpeng@ustc.edu.cn) or L. Y. H. (Lyhung@fudan.edu.cn) or Y. D. W. (ydwan@fudan.edu.cn).
\end{addendum}

\clearpage

\clearpage

\clearpage

\clearpage

\clearpage

\clearpage
\setcounter{figure}{0}
\setcounter{equation}{0}
\renewcommand{\thefigure}{S\arabic{figure}}
\renewcommand{\theequation}{S\arabic{equation}}
\begin{center}
{\bf{SUPPLEMENTARY INFORMATION}\\
\bf{for}\\
\bf{Experimentally Probing Topological Order and its breakdown via Modular Matrices}\\}
\author{Zhihuang Luo$^{1,2,3}$, Jun Li$^{2}$, Zhaokai Li$^{1}$, Ling-Yan Hung${^{4,5,6}}^{*}$, Yidun Wan${^{5,6,7}}^{*}$, Xinhua Peng${^{1,8}}^{*}$, \& Jiangfeng Du$^{1,8}$}
\end{center}

\section{The modular $S$ and $T$ matrices}

The topological properties of $2+1$-dimensional quantum systems with an energy gap can be described by topological quantum field theories (TQFTs) or unitary modular tensor category \cite{Kitaev2006,Wang2014}. We begin with a finite set $\mathcal{C}$ of quasiparticle or anyon types labeled by $a,b,c,\cdots\in\mathcal{C}$. The two major concepts in anyon models are fusion and braiding, which can be represented diagrammatically by oriented, labeled particle worldlines, and are unaffected by smooth deformations in which the lines do not intersect. The fusion rules are $a\times b=\sum_{c\in\mathcal{C}}N_{ab}^cc$, where the integer $N_{ab}^c$ is the dimension of the Hilbert space of particles of type a and b restricted to have total anyonic charge $c$. The braiding operator of $a$ and $b$ is represented diagrammatically by
\begin{equation}\label{}
    \tikz[baseline=-3,thick]{\draw (0,0)--(0,-0.3) [<-,>=latex] (0,-0.2)--(0,-0.6);\draw[] (0,0)arc[start angle=-60,end angle=35,radius=0.2];\draw (0,0)arc[start angle=-120,end angle=-240,radius=0.2];\draw (0,0.3464)--(0.3464,0.5464);\draw[->,>=latex] (0,0.3464)--(0.2598,0.4964);\draw (-0.0433,0.3714)--(-0.3464,0.5464);\draw[->,>=latex] (-0.0433,0.3714)--(-0.2598,0.4964);\node[left] at(0,-0.4){\footnotesize c};\node[right] at(0,0){\footnotesize $\mu$};\node[left] at(-0.35,0.35){\footnotesize a};\node[right] at(0.35,0.35){\footnotesize b};}=\sum_{\nu}\left[R_c^{ab}\right]_{\mu\nu} \tikz[baseline=-3,thick]{\draw (0,0)--(0,-0.3) [<-,>=latex] (0,-0.2)--(0,-0.6);\draw (0.2121,0.2121)--(0.4242,0.4242) [->,>=latex] (0,0)--(0.35,0.35);\draw (-0.2121,0.2121)--(-0.4242,0.4242) [->,>=latex] (0,0)--(-0.35,0.35);\node[right] at(0,0){\footnotesize $\nu$};\node[left] at(-0.35,0.35){\footnotesize a};\node[right] at(0.35,0.35){\footnotesize b};\node[left] at(0,-0.4){\footnotesize c};}.
\end{equation}
The above finite set $\mathcal{C}$, fusion rules $N_{ab}^c$， and braiding rules $R_c^{ab}$ uniquely determine an anyon model. The other physical quantities or properties can be derived from these three definitions. For example, according to braiding, we can define an important quantity---topological spin as,
\begin{equation}\label{}
    \theta_a=\theta_{\bar{a}}=\sum_{c,\mu}\frac{d_c}{d_a}\left[R_c^{aa}\right]_{\mu\mu}=\frac{1}{d_a}\tikz[baseline=-3,thick]{\draw (0,0)--(0.2828,0.2828) [->,>=latex] (-0.2828,-0.2828)--(0,0);\draw (-0.2828,0.2828)--(-0.08,0.08);\draw (0.2828,-0.2828)--(0.08,-0.08);\draw (0.2828,0.2828)arc[start angle=135,end angle=-135,radius=0.4];\draw (-0.2828,-0.2828)arc[start angle=-45,end angle=-315,radius=0.4];\node[below] at(0,-0.1){a};}.
\end{equation}
We now give the algebraic definitions of the modular $S$ and $T$ matrices:
\begin{equation}\label{}
    S_{ab}=\mathcal{D}^{-1}\sum_c N_{\bar{a}b}^c \frac{\theta_c}{\theta_a\theta_b}d_c=\frac{1}{\mathcal{D}}\tikz[baseline=-3]{\draw[thick] (0.6,0.3464)arc[start angle=60,end angle=390,radius=0.4];\draw[thick] (0.6193,-0.2)arc[start angle=210,end angle=-120,radius=0.4];\draw[thick,<-,>=latex] (0,0.1)arc[start angle=180,end angle=181,radius=0.4];\draw[thick,<-,>=latex] (0.5657,0.1)arc[start angle=180,end angle=181,radius=0.4];\node[below left] at(0,0){a};\node[below left] at(0.5657,0.1){b};},
\end{equation}
and $T_{ab}=\theta_{a}\delta_{ab}$, where $\mathcal{D}=\sqrt{\sum_ad_a^2}$ is total quantum dimension, and $d_a$  is the quantum dimension of anyon $a$ that takes the value of a single loop of that type, namely,
\begin{equation}\label{}
    d_a=d_{\bar{a}}=\tikz[baseline=-3]{\draw[thick,<-,>=latex] (0,0)arc[start angle=-185,end angle=180,radius=0.4] node[below left]{a};}.
\end{equation}

From the above algebraic theory of anyons or unitary modular tensor category, it is not difficult to find that the elements of a modular $S$ matrix determines the mutual statistics of anyons, while the elements of a $T$ matrix record the topological spins. We can further reconstruct the fusion coefficients from an $S$ matrix according to the Verlinde formula, i.e.,
\begin{equation}\label{}
    N_{ab}^c=\sum_{x\in\mathcal{C}}\frac{S_{ax}S_{bx}S_{cx}^{*}}{S_{0x}}.
\end{equation}
The fusion rules and braiding rules reflect two major algebraic structures of anyons. The topological properties of topologically ordered states are associated to elementary excitations, i.e., anyons. Therefore, $S$ and $T$ matrices provide a complete description and can be taken as the nonlocal order parameters of topological orders.

To better understand this point, we list several examples of different topological orders in Table \ref{tab:ST}\cite{Liu2014v2}. The numerical results show that the modular matrices (up to a unitary transformation) furnish a complete and one-to-one characterization of exact topological orders.

\begin{table}
\caption{The anyon types and the standard modular matrices for $\mathbb{Z}_2$ Toric code order, doubled semion order and doubled fibonacci order.}
\label{tab:ST}
%\footnotesize
\centering
\begin{tabular}{|c||c|c|}
  \hline
  Orders & Anyon types & Standard modular matrices \\ \hline\hline
  $\mathbb{Z}_2$ Toric code & $1, e, m, f$ & $\begin{aligned}
  S=\frac{1}{2}&\left(
    \begin{array}{cccc}
    1 & 1 & 1 & 1\\
    1 & 1 & -1 & -1\\
    1 & -1 & 1 & -1\\
    1 & -1 & -1 & 1
    \end{array}
    \right), \\
T&=\text{diag}\{1,1,1,-1\}
\end{aligned}$ \\ \hline
  Doubled semion & $1, s, \bar{s}, s\bar{s}$ & $\begin{aligned}
  S=\frac{1}{\sqrt{2}}&\left(
    \begin{array}{cc}
    1 & 1 \\
    1 & -1
    \end{array}
    \right)\bigotimes\frac{1}{\sqrt{2}}\left(
    \begin{array}{cc}
    1 & 1 \\
    1 & -1
    \end{array}
    \right), \\
T&=\text{diag}\{1,i\}\bigotimes\text{diag}\{1,-i\}
\end{aligned}$ \\ \hline
  Doubled fibonacci & $1, \tau, \bar{\tau}, \tau\bar{\tau}$ &  $\begin{aligned}
  S=\frac{5-\sqrt{5}}{\sqrt{10}}&\left(
    \begin{array}{cc}
    1 & \frac{1+\sqrt{5}}{2} \\
    \frac{1+\sqrt{5}}{2} & -1
    \end{array}
    \right)\bigotimes\left(
    \begin{array}{cc}
    1 & \frac{1+\sqrt{5}}{2} \\
    \frac{1+\sqrt{5}}{2} & -1
    \end{array}
    \right), \\
T&=\text{diag}\{1,e^{-i\frac{4\pi}{5}}\}\bigotimes\text{diag}\{1,e^{i\frac{4\pi}{5}}\}
\end{aligned}$ \\
  \hline
\end{tabular}
\end{table}

\section{Comparing small and large systems}

In the main text, we discussed the identification of a phase transition point. Attention is needed since phase transitions are only properly defined in the thermodynamic limit. We have to adjust the notion such that it is still reasonable in such a small system, and also that it asymptotes to the true phase transition in the thermodynamic limit. In addition, there is another issue that plagues small systems: the ground state degeneracy is generically far from being exact, except at the special Kitaev solvable point. This is so because the topological degeneracy is protected only in the thermodynamic limit, with a splitting that scales like $\exp(N)$, for lattice size $N$.  When we deviate from the exactly solvable point at finite $h$ and $\epsilon_i$, the four degenerate ground states acquire a finite splitting. There remains however a visible gap between the bottom four states continuously connected to the originally degenerate ground states and the rest of the excited states. As $h$ is varied, the higher states eventually cross the lowest four states. We will therefore define the phase transition point to be located at  $h_c$ , at which any one or more of the higher states first cross any of the lowest four states.
We note that in the thermodynamic limit, phase transitions can  also be characterized by the closing of the topological energy gap. Therefore our definition is the closest physical analogue of phase transitions in a small system that by construction asymptotes to the usual definition in the thermodynamic limit.

In Fig. \ref{fig:gap_mag_SM} below, we show a plot of the energy gap between the 5th excited state and the ground state sub-space (the splitting between the ground states are very small in comparison) and the magnetization of the ground state and the 5th excited state as the detuning parameter $h$ is varied.  These can be compared with Fig. 2-3 in \cite{Nayak2007} of the same model in the thermodynamic limit. The closing of the gap shows similarity. For the magnetization, the transition in the thermodynamic limit is continuous, and so the magnetization is continuous across transition. In the 4-spin system, there is only a level crossing and the magnetizations of the two states are different. We note however that they increases as the magnetic coupling $h$ increases, as expected.

\begin{figure}
\centering
\begin{tabular}{c c}
 \includegraphics[width=7cm]{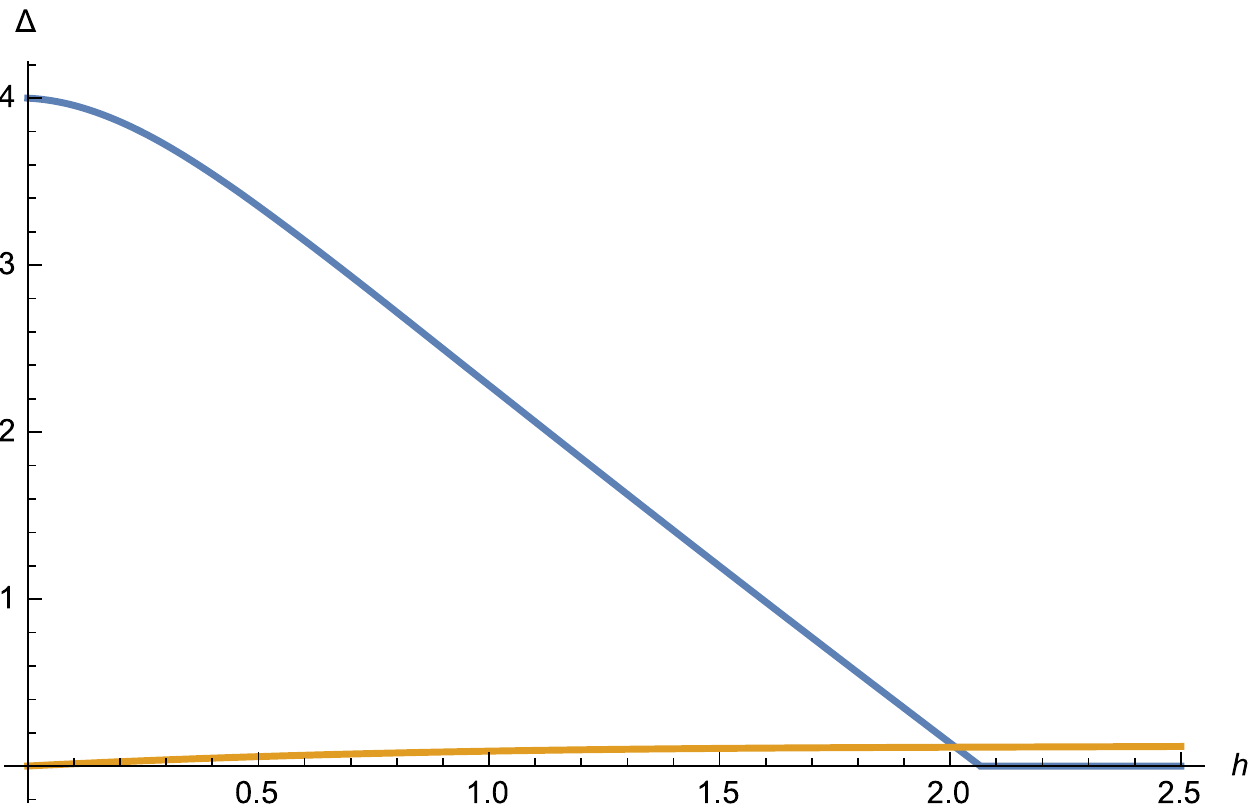} & \includegraphics[width=7cm]{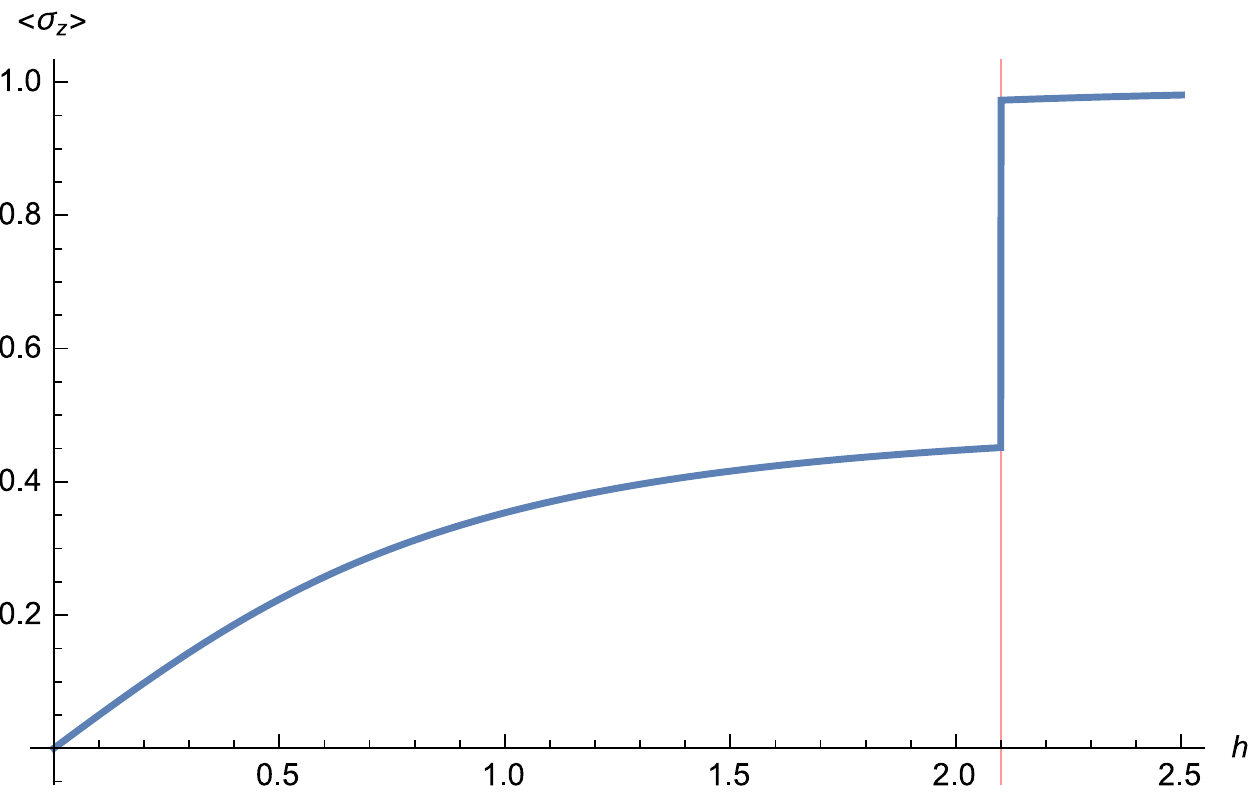} \\
 (a) & (b)
 \end{tabular}
  \caption{(a) (Blue curve ) Energy gap between the ground state and the fifth-excited state.  (Orange curve) The splitting between the ground state subspace. (b) Magnetization of the ground state. $h_c$ is indicated by the red line and there is a sharp jump since the excited state, with a different magnetization, has taken the place of the ground state at level crossing at $h_c$ (indicated by the red vertical line). }\label{fig:gap_mag_SM}
\end{figure}

\begin{figure}
  \centering
 \includegraphics[width=8cm]{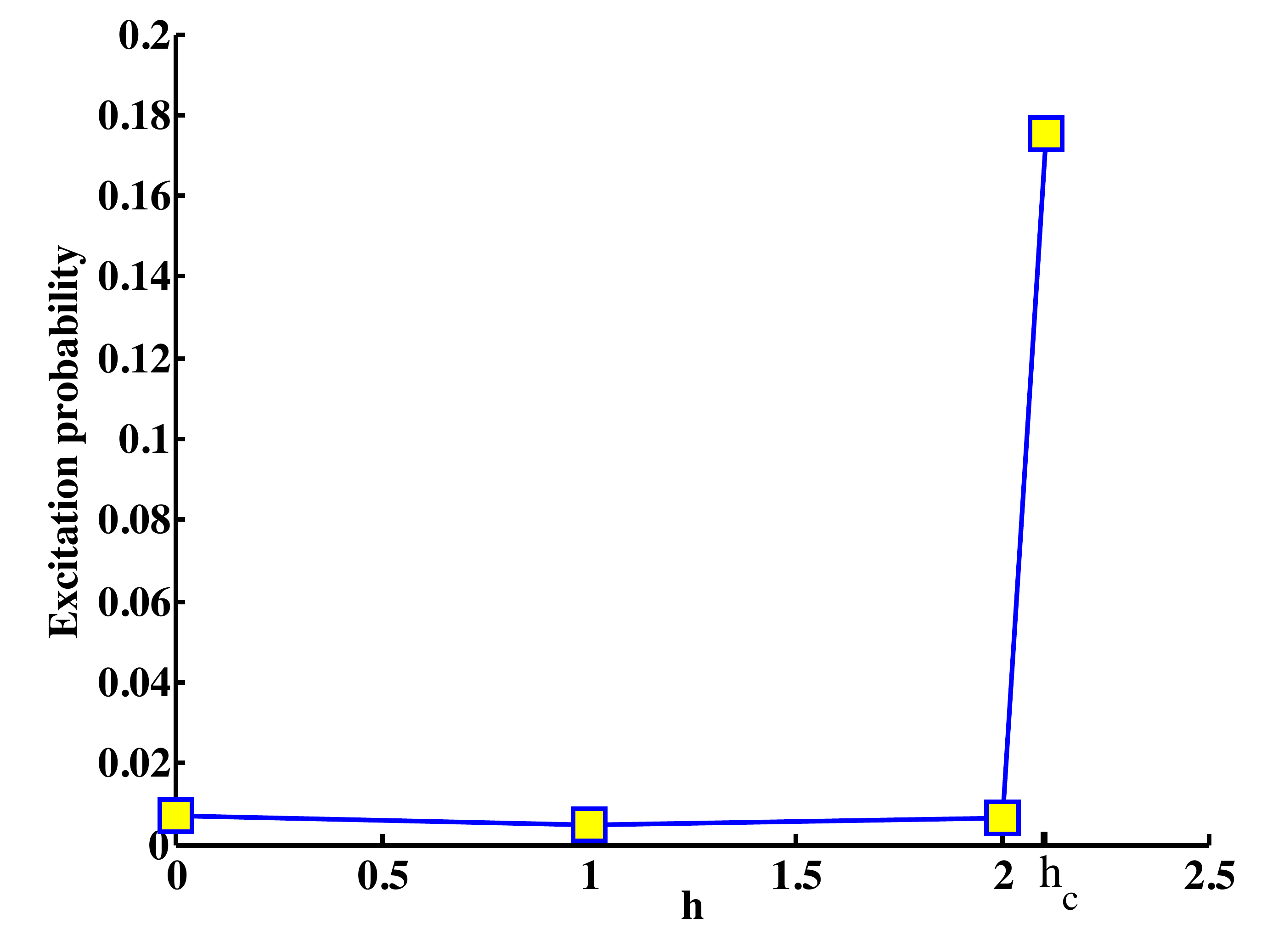}
  \caption{Excitation probability beyond the ground state subspace at fixed adiabatic time scale $T$ given in the supplementary information as the detuning parameter $h$ varies. The excitation probability shows a sharp increase as $h_c$ is approached.}\label{fig:prob_SM}
\end{figure}

As shown in Fig. \ref{fig:gap_mag_SM} , it is evident that as  we zoom into the transition point, the topological gap closes, and one would expect that as we go sufficiently close to $h_c$, the finite size splitting becomes comparable to the topological gap, and the adiabatic method that we adopt for ground state preparation would potentially  break down, leading to significant excitation beyond the ground state subspace. We show in Fig. \ref{fig:prob_SM} the excitation probability for the adiabatic time scale we have adopted as explained in detail below. This is to be compared with the thermal excitation probability discussed in figure 3 of \cite{Nayak2007}. There, it shows an even sharper increase as the detuning parameter $h$ approaches phase transition in the thermodynamic limit than we have in Fig. \ref{fig:prob_SM}. It is evident that the excitation probability remain small even as we get very close to $h_c$. Such a feature would improve the experimental accuracy determining the location of the transition, allowing one to get closer to $h_c$ before the adiabatic method breaks down.

\section{Experimental operations for implementing modular $S$ and $T$ transformations}
\begin{figure}
  \centering
  \includegraphics[width=13cm]{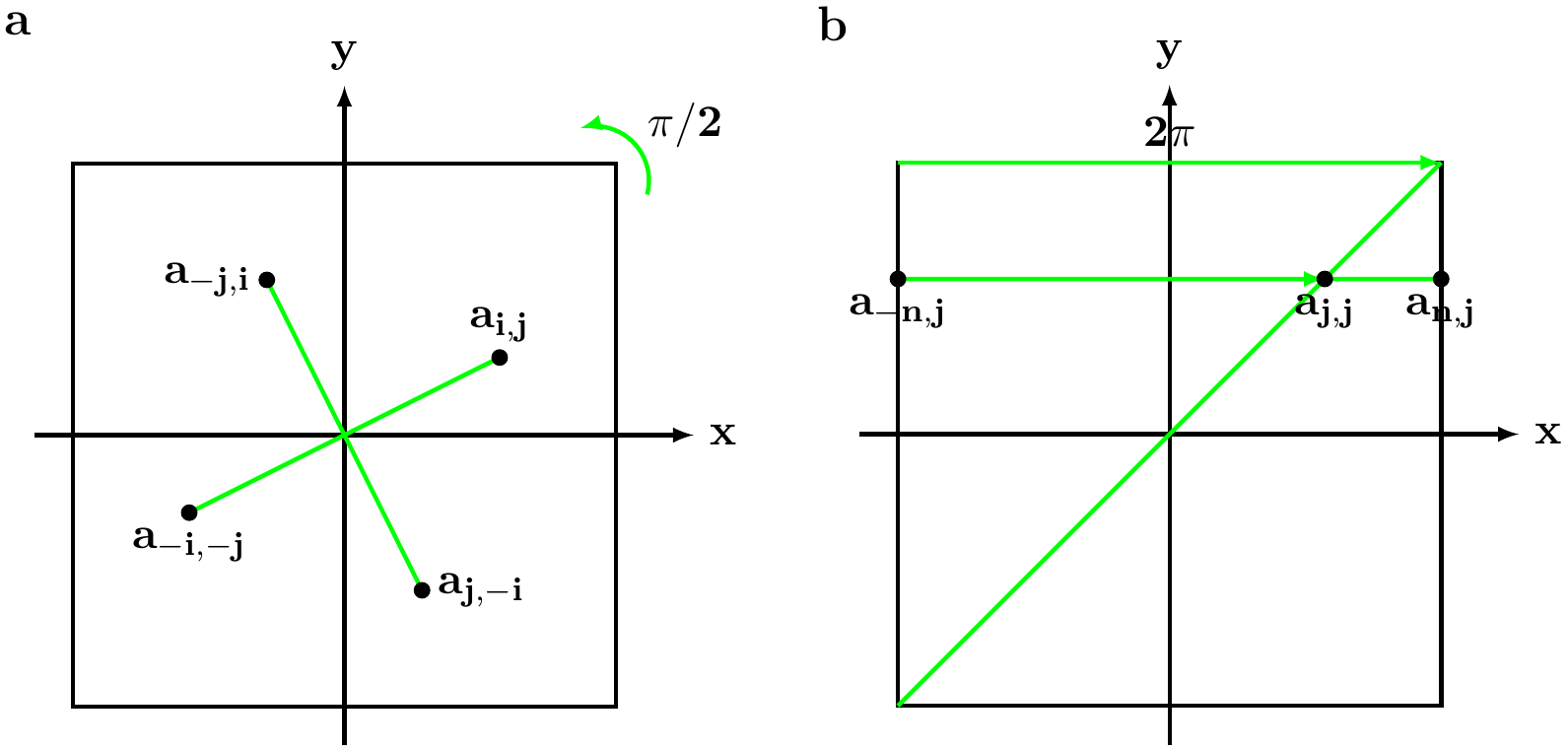}\\
  \caption{Diagrams for implementing a $\pi/2$ rotation and a shear, respectively.}\label{fig:1_SM}
\end{figure}

For an $N\times N$ square lattice of the Kitaev toric code model, the implementation of the modular $S$ and $T$ transformations can be realized by a sequence of SWAP operations in experiments. Here we give a general procedure for their experimental operations and further analyze their complexities. Because the modular $S$ and $T$ transformations correspond respectively to a $\pi/2$ rotation and a shear, they are equivalent to a series of permutations with length $4$ and $N$, respectively, as illustrated in Figs. \ref{fig:1_SM}a and \ref{fig:1_SM}b. For example, four associated points in Fig. \ref{fig:1_SM}a form a cycle under a $\pi/2$ rotation, i.e., $(a_{i,j},a_{j,-i},a_{-i,-j},a_{-j,i})$
, and $N$ associated points $(a_{-n,j}=a_{n,j}, n=\lfloor(N+1)/2\rfloor)$ in Fig. \ref{fig:1_SM}b form a permutation under a shear, i.e., $\left(\begin{array}{ccccc}
    a_{-n,j} & a_{-n+1,j} & \cdots & a_{n-2,j} & a_{n-1,j} \\
    a_{j,j} & a_{j+1,j} & \cdots & a_{j-2,j} & a_{j-1,j}
    \end{array}\right)$,
which can be decomposed as the product of $n_m(=\text{gcd}(n+j,N)$, where gcd means the greatest common divisor) cycles of length $m=N/n_m$. According to group theory, any cycle can be written as the product of transpositions or SWAP operations between two elements. We thus have
\begin{equation}\label{}
   (a_{i,j},a_{j,-i},a_{-i,-j},a_{-j,i})=(a_{i,j},a_{-j,i})(a_{i,j},a_{-i,-j})(a_{i,j},a_{j,-i}),
\end{equation}
and
\begin{align}\label{}
    &\left(\begin{array}{ccccc}
    a_{-n,j} & a_{-n+1,j} & \cdots & a_{n-2,j} & a_{n-1,j} \\
    a_{j,j} & a_{j+1,j} & \cdots & a_{j-2,j} & a_{j-1,j}
    \end{array}\right)  \\ \nonumber
   & = \prod_{k=0}^{n_m-1}(a_{-n+k,j},a_{j+k,j},\cdots,a_{(m-2)(n+j)-n+k \text{ mod } N,j},a_{(m-1)(n+j)-n+k \text{ mod } N,j}) \\ \nonumber
   &=\prod_{k=0}^{n_m-1}(a_{-n+k,j},a_{(m-1)(n+j)-n+k \text{ mod } N,j})(a_{-n+k,j},a_{(m-2)(n+j)-n+k \text{ mod } N,j})\cdots(a_{-n+k,j},a_{j+k,j}).
\end{align}

We now can calculate the number of experimental operations. For implementing S transformation, it needs $\#(\text{SWAP})=3(n^2-1)+3(n-1)+1=3n^2+3n+5$. Here we consider the situation on the border separately, because it is different a little from that in the body of lattice. For implementing T transformation, it needs $\#(\text{SWAP})=\sum_{j=-n+1}^{n-1}(m-1)n_m=N^2-N-\sum_{j=-n+1}^{n-1}\text{gcd}(n+j,N)$. So the implementations of modular $S$ and $T$ transformations only require polynomial SWAP gates.

\section{Decomposition of controlled modular matrix}

\begin{figure}
  \centering
  \includegraphics[width=7.5cm]{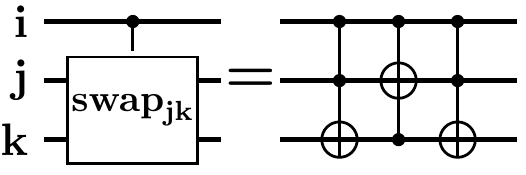}\\
  \caption{Decomposition of the controlled $S$ or $T$ matrix for our experimental proposal.}\label{fig:2_SM}
\end{figure}

Because the modular $S$ and $T$ transformations are implemented experimentally by the product of a series of SWAP operations. And a controlled-$U$ gate is equivalent to the product of a controlled-$B$ gate and a controlled-$A$ gate, if $U$ satisfies that $U=AB$ \cite{Nielsen2000}. Therefore, the controlled modular matrices can be decomposed into a series of Toffoli gates. For instance, in our experimental proposal (i.e., only considering a $2\times2$ lattice), the decomposition of controlled modular $S$ or $T$ matrix are shown in Figs. \ref{fig:2_SM}. It is not difficult to see that it requires $3\#(\text{SWAP})$ Toffoli gates to implement a controlled modular matrix for a $N\times N$ lattice. The decomposition complexity is polynomial, which means that the measurement of the modular $S$ and $T$ matrices is an efficient approach for large systems.

\section{Experimental details and results}

\begin{figure}
  \centering
  \includegraphics[width=16cm]{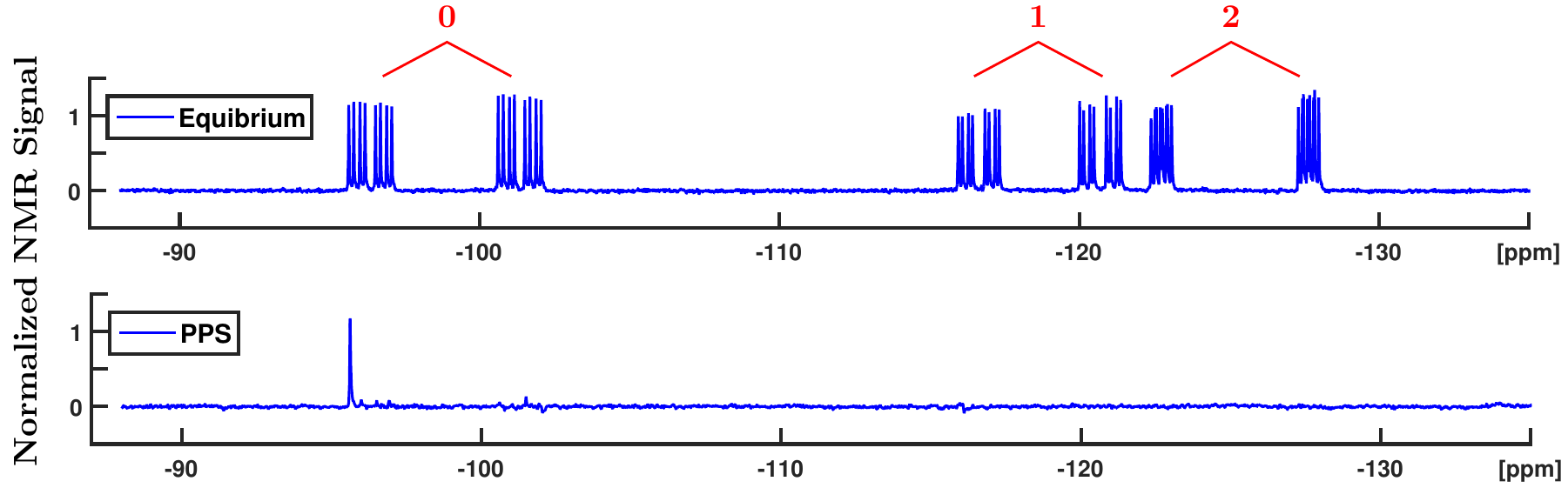}\\
  \caption{Experimental $^{19}$F spectra of thermal equilibrium state (at the top) and PPS $\rho_{00000}$ (at the bottom). }\label{fig:3_SM}
\end{figure}

The equilibrium $^{19}$F spectrum of 1-bromo-2,4,5-trifluorobenzene is illustrated at the top of Fig. \ref{fig:3_SM}. The first $^{19}$F nuclear spin (labeled by 0) is used for interference measurement. The spectrum of PPS $\rho_{00000}$ is shown at the bottom of Fig. \ref{fig:3_SM}, which was prepared from the thermal equilibrium state $\rho_{\text{eq}}$ by two line-selective shaped pulses. The first shaped pulse with length of 37.5 ms was applied to selectively excite the energy levels and make the populations equal of all levels except for $00000$-level. It can be written as the product of a sequence of line-selective pulses, i.e.,
\begin{equation}\label{}
    U_1=\prod_{n=1}^{30}e^{-i\theta_n\hat{I}_x^{(n_1,n_2)}},
\end{equation}
where $\hat{I}_x^{(n_1,n_2)}$ is the single-transition operator between energy levels $n_1$ and $n_2$. The flipping angles $\{\theta_n\}_{n=1,2,\cdots,30}$ were optimized by numerical search procedure, with appropriate values that satisfies the following condition: $\text{diag}[U_1\rho_{\text{eq}}U_1^{\dag}]=\text{diag}[\rho_{00000}]$.
The second shaped pulse with length of 15 ms was applied to remove zero quantum coherence that cannot be averaged out by $z$-direction gradient fields. It consists of several controlled-NOT gates between qubits $n$ and $m$, denoted as $\text{CNOT}_{nm}$. For example, the second shaped pulse $U_2=\text{CNOT}_{12}\text{CNOT}_{24}\text{CNOT}_{51}$ was chosen in our experiment. The intensity ratio for the leftmost peak in PPS $\rho_{00000}$ and the reference in the equilibrium spectrum is $1.12$. The signal loss is mainly due to relaxation effect and gradient fields.

\begin{figure}
  \centering
  \includegraphics[width=16cm]{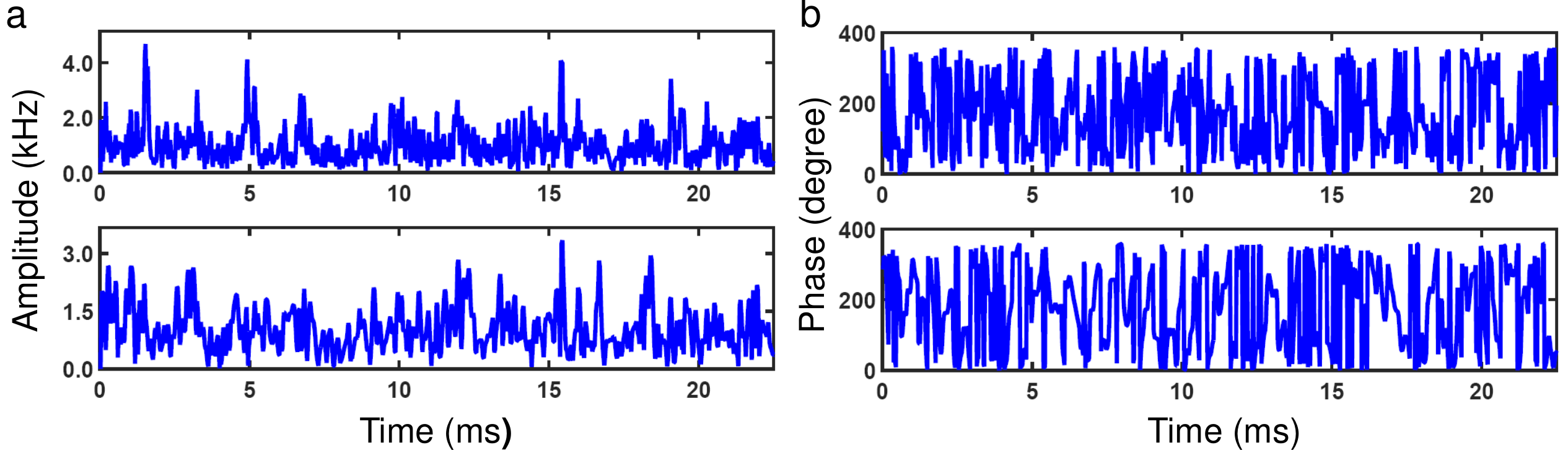}\\
  \caption{The shaped pulse for implementing one adiabatic passage. {\textbf{a,}} The amplitude and {\textbf{b,}} the phase as the functions of time,  applying for the $^{19}$F (top) and $^{1}$H (bottom) channels respectively.}\label{fig:waveform}
\end{figure}

In experiments, \emph{random adiabatic evolution} was discretized $M=100$ steps and the duration of each step is $\tau\approx 2.82$. The Hamiltonian of each step can be efficiently simulated using the radio-frequency (RF) pulses and two-body interactions in the NMR system \cite{Luo2014}. The details of adiabatic evolution can been see in Ref.\cite{Luo2016}. To reduce the accumulated pulse errors in experiments, the shaped pulses calculated by the gradient ascent pulse engineering (GRAPE) method \cite{Glaser2005} are applied to the adiabatic passages, with the length of each pulse being 22.5 ms shown in Fig. \ref{fig:waveform}. All the pulses have theoretical fidelities over 0.99, and are designed to be robust against the inhomogeneity of radio-frequency pulses.

The experimental spectra for measuring the overlaps of four randomly generated linearly independent ground states are shown in Figs. \ref{fig:6_SM}a $\sim$ \ref{fig:6_SM}d, corresponding to four different magnetic fields: $h=0,h=1,h=2,$ and $h=2\sqrt{10}/3$, respectively. The values of overlaps $\langle\psi_i^{\text{rd}}|\psi_j^{\text{rd}}\rangle, i<j\in [2,4]$ are listed in Table \ref{tab:overlap}. The coefficients expanded in the orthogonal basis (i.e., $a_j^i=\langle\phi_i|\psi_j^{\text{rd}}\rangle$) can therefore be solved from the above overlaps, as shown in Table \ref{tab:coefficient}.

\begin{table}
\caption{The experimental overlaps of $\langle\psi_i^{\text{rd}}|\psi_j^{\text{rd}}\rangle$s in different magnetic fields $h$.}
\label{tab:overlap}
%\footnotesize
\centering
\begin{tabular}{|c||c|c|c|c|}
  \hline
   & $h=0$ & $h=1$ & $h=2$ & $h=2\sqrt{10}/3$ \\ \hline\hline
  $\langle\psi_1^{\text{rd}}|\psi_2^{\text{rd}}\rangle$ & $-0.3344 + 0.4432i$ & $0.3206 - 0.5970i$ & $0.5215 - 0.0982i$ & 0.5357 - 0.0288i \\
  $\langle\psi_1^{\text{rd}}|\psi_3^{\text{rd}}\rangle$ & $0.6932 - 0.1028i$ & $0.0466 + 0.0958i$ & $0.3355 + 0.6402i$ & $0.0724 + 0.4777i$ \\
  $\langle\psi_1^{\text{rd}}|\psi_4^{\text{rd}}\rangle$ & $0.1882 + 0.1908i$ & $-0.5954 - 0.2380i$ & $-0.3913 - 0.0048i$ & $ 0.1789 - 0.4556i$ \\
  $\langle\psi_2^{\text{rd}}|\psi_3^{\text{rd}}\rangle$ & $-0.2362 - 0.4322i$ & $0.1922 - 0.4100i$ & $-0.0355 + 0.2212i$ & $0.0466 + 0.7724i$ \\
  $\langle\psi_2^{\text{rd}}|\psi_4^{\text{rd}}\rangle$ & $0.1532 - 0.0482i$ & $-0.0448 - 0.3286i$ & $-0.1691 - 0.1220i$ & $0.2245 - 0.6482i$ \\
  $\langle\psi_3^{\text{rd}}|\psi_4^{\text{rd}}\rangle$ & $ -0.0942 + 0.4642i$ & $-0.1256 + 0.0472i$ & $-0.3854 + 0.6774i$ & $ -0.6048 - 0.2110i$ \\
  \hline
\end{tabular}
\end{table}

\begin{table}
\caption{The coefficients of $a_j^i$s obtained from the experimental overlaps.}
\label{tab:coefficient}
%\footnotesize
\centering
\begin{tabular}{|c||c|c|c|c|}
  \hline
   & $h=0$ & $h=1$ & $h=2$ & $h=2\sqrt{10}/3$ \\ \hline\hline
  $a_2^1$ & $-0.3344 + 0.4432i$ & $0.3206 - 0.5970i$ & $0.5215 - 0.0982i$ & 0.5357 - 0.0288i \\
  $a_2^2$ & $0.8317 + 0.0000i$ & $0.7354 + 0.0000i$ & $0.8476 + 0.0000i$ & $0.8439 + 0.0000i$ \\ \hline
  $a_3^1$ & $0.6932 - 0.1028i$ & $0.0466 + 0.0958i$ & $0.3355 + 0.6402i$ & $ 0.0724 + 0.4777i$ \\
  $a_3^2$ & $0.0495 - 0.1916i$ & $0.3188 - 0.6371i$ & $-0.1742 - 0.1719i$ & $0.0256 + 0.6095i$ \\
  $a_3^3$ & $0.6854 + 0.0000i$ & $0.6936 + 0.0000i$ & $0.6462 + 0.0000i$ & $0.6280 + 0.0000i$ \\ \hline
  $a_4^1$ & $0.1882 + 0.1908i$ & $-0.5954 - 0.2380i$ & $-0.3913 - 0.0048i$ & $ 0.1789 - 0.4556i$ \\
  $a_4^2$ & $0.1582 + 0.1190i$ & $0.0054 + 0.1403i$ & $0.0408 - 0.0956i$ & $ 0.1369 - 0.4850i$ \\
  $a_4^3$ & $-0.2773 + 0.4033i$ & $0.0181 - 0.0677i$ & $-0.4028 + 0.6264i$ & $-0.1720 + 0.0053i$ \\
  $a_4^4$ & $0.8059 + 0.0000i$ & $0.7512 + 0.0000i$ & $0.5304 + 0.0000i$ & $ 0.6906 + 0.0000i$ \\  \hline
\end{tabular}
\end{table}

\begin{figure}[!htp]
  \centering
  \includegraphics[width=17cm]{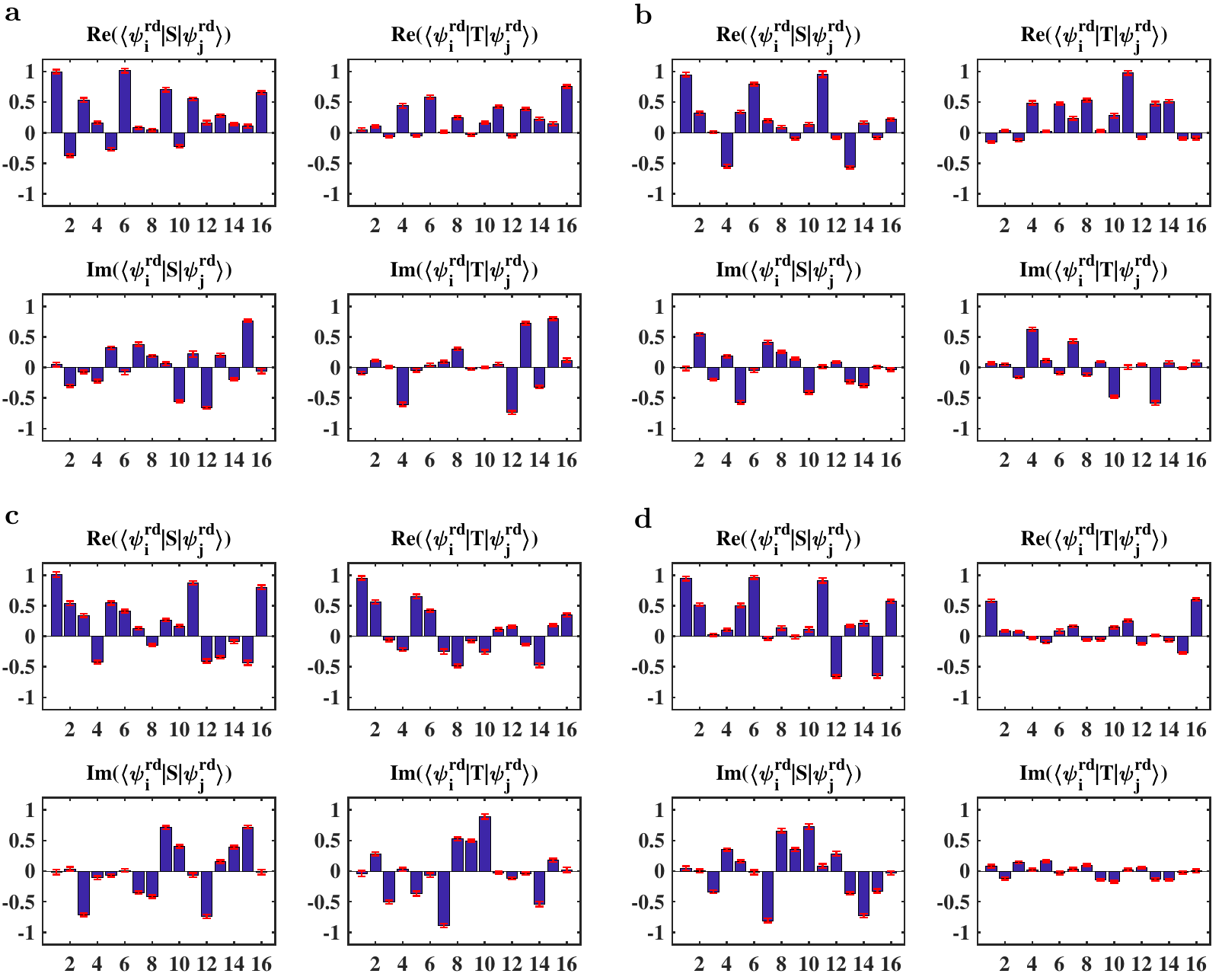}\\
  \caption{Experimental $S$, $T$ matrices in randomly generated linearly independent ground states $\{|\psi_i^{\text{rd}}\rangle\}$ for different magnetic fields: {\textbf{a,}} $h=0$; {\text{b,}} $h=1$; {\textbf{c,}} $h=2$; {\textbf{d}}, $h=2\sqrt{10}/3$.}\label{fig:11_SM}
\end{figure}

\begin{figure}[!htp]
  \centering
  \includegraphics[width=17cm]{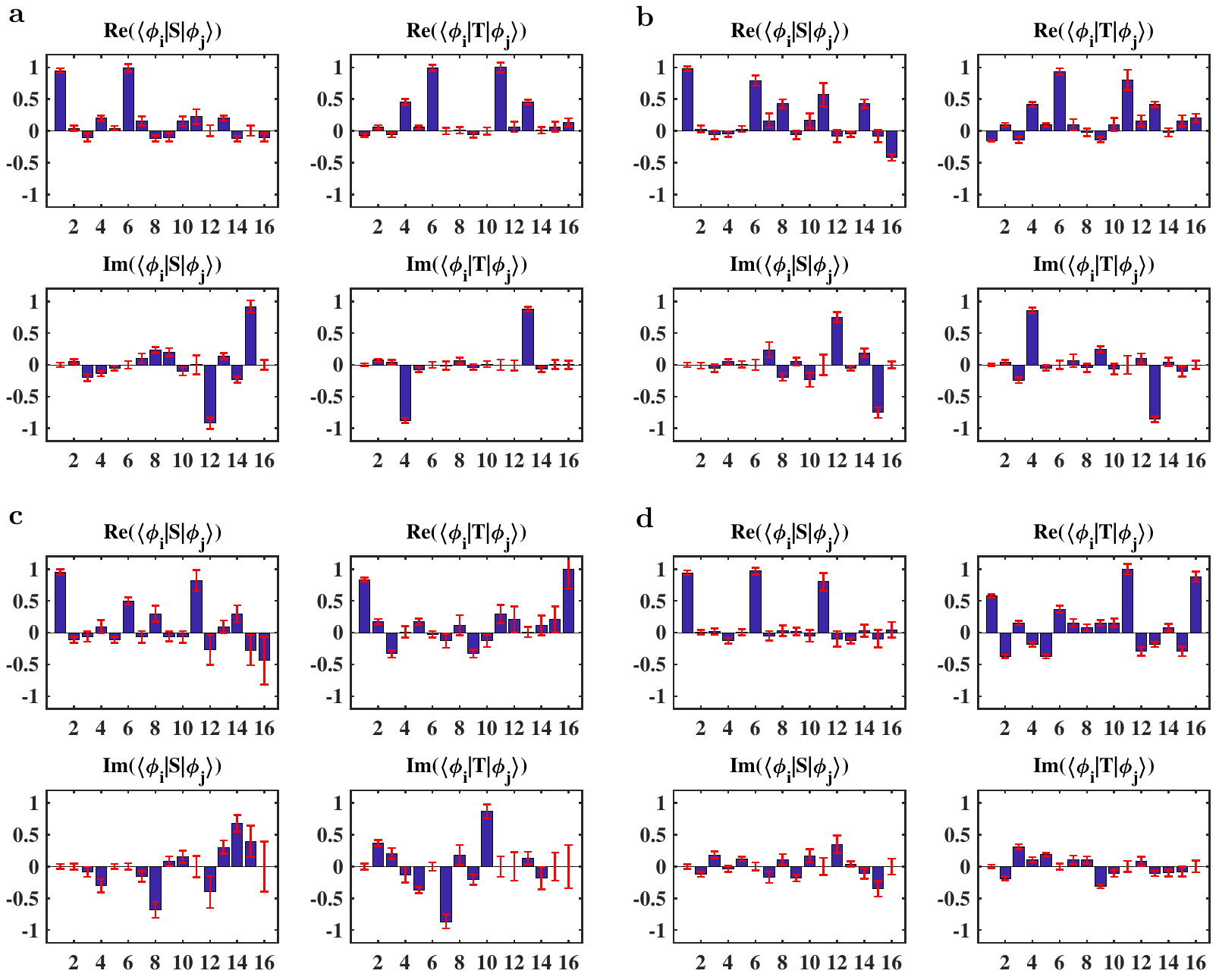}\\
  \caption{Resulting $S$ and $T$ matrices in the orthogonal basis $\{|\phi_i\rangle\}$ for different magnetic fields: {\textbf{a,}} $h=0$; {\text{b,}} $h=1$; {\textbf{c,}} $h=2$; {\textbf{d}}, $h=2\sqrt{10}/3$.}\label{fig:12_SM}
\end{figure}

\begin{figure}[!htp]
  \centering
  \includegraphics[width=17cm]{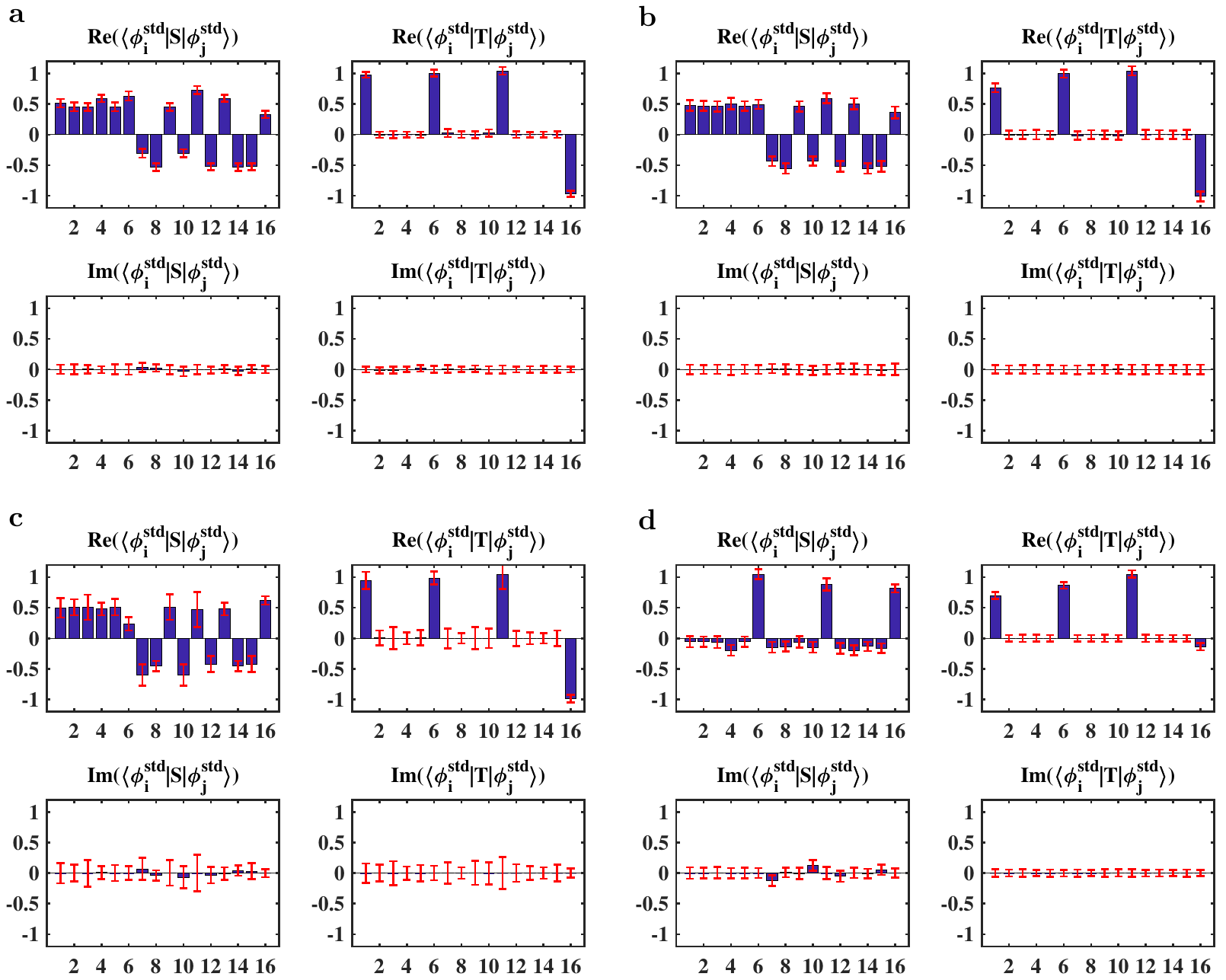}\\
  \caption{Resulting $S$ and $T$ matrices in the standard basis $\{|\phi_i^{\text{std}}\rangle\}$ for different magnetic fields: {\textbf{a,}} $h=0$; {\text{b,}} $h=1$; {\textbf{c,}} $h=2$; {\textbf{d}}, $h=2\sqrt{10}/3$.}\label{fig:13_SM}
\end{figure}

\begin{figure}[!htp]
  \centering
  \includegraphics[width=17cm]{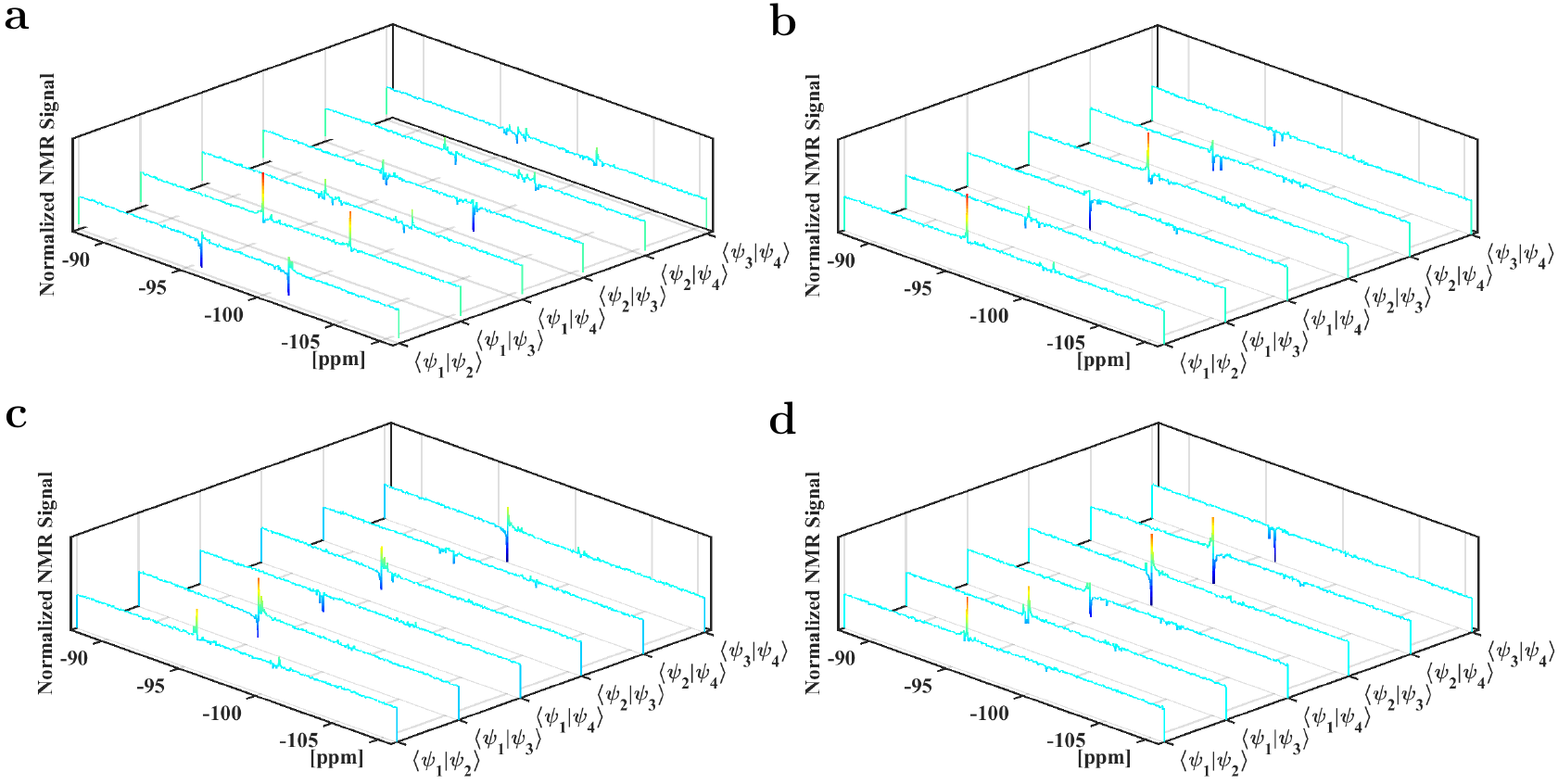}\\
  \caption{Experimental $^{19}$F spectra for measuring the overlaps of randomly generated linearly independent ground states $\{|\psi_i^{\text{rd}}\rangle\}$ in the different magnetic fields: {\textbf{a,}} $h=0$; {\text{b,}} $h=1$; {\textbf{c,}} $h=2$; {\textbf{d}}, $h=2\sqrt{10}/3$.}\label{fig:6_SM}
\end{figure}

\begin{figure}[!htp]
  \centering
  \includegraphics[width=17cm]{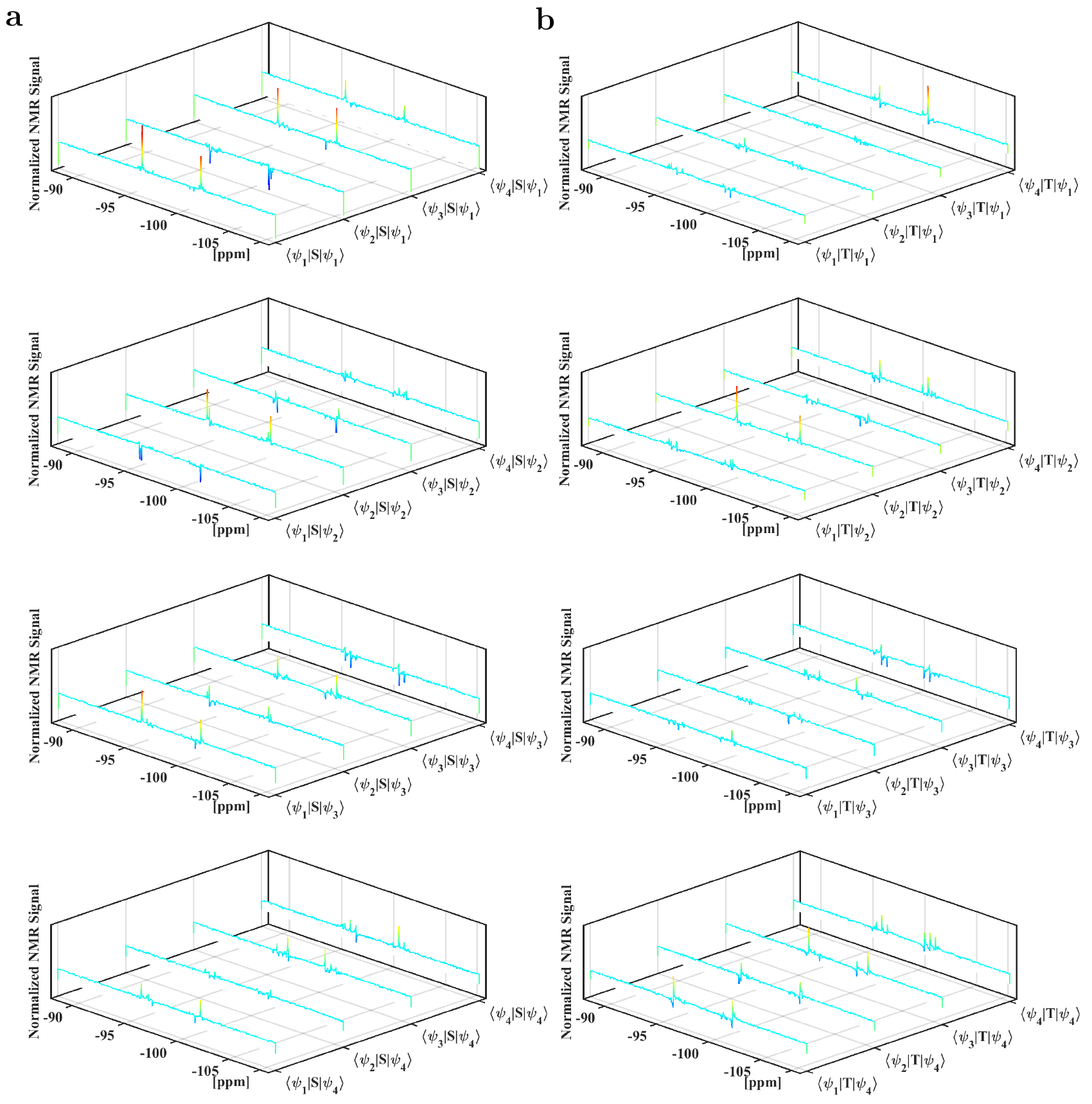}\\
  \caption{Experimental $^{19}$F spectra of measuring all elements of $S$ and $T$ matrices in the random basis $\{|\psi_i^{\text{rd}}\rangle\}$ for $h=0$.}\label{fig:7_SM}
\end{figure}

Figures \ref{fig:7_SM} $\sim$ \ref{fig:10_SM} show the $^{19}$F spectra of measuring all elements of $S$, $T$ matrices in four randomly generated ground states for four different magnetic fields: $h=0,h=1,h=2,$ and $h=2\sqrt{10}/3$, respectively. The total integrals of all peaks in each spectrum represent the element value of the matrices. The experimental $S$ and $T$ matrices in the random basis are illustrated in Fig. \ref{fig:11_SM}, where the error bars come from the line shape fitting procedure. The $S$ and $T$ matrices in the orthogonal basis can be obtained by applying the transformation $A$ (see equation (8) in Method) on the $S$ and $T$ matrices in the random basis. The results are plotted in Fig. \ref{fig:12_SM}. Following the recovering algorithm (see Method), we finally achieve the standard $S$ and $T$ matrices shown in Fig. \ref{fig:13_SM}. The error bars in Figs. \ref{fig:12_SM} and \ref{fig:13_SM} are slightly amplified due to the error transfer in the basis transformations.

%\clearpage
\begin{center}
  \rotatebox{90}{%
    $
A=\left(
    \begin{array}{cccccccccccccccc}
     1  &   0   &  0  &   0  &   0   &  0   &  0  &   0   &  0  &   0  &   0  &   0  &   0   &  0  &   0  &   0 \\
     a_2^{1*} &    a_2^{2*} &    0  &   0  &   0  &   0  &   0  &   0  &   0  &   0  &   0  &   0  &   0   &  0  &   0   &  0 \\
     a_3^{1*}  &   a_3^{2*}  &   a_3^{3*}  &   0   &  0 &    0  &   0   &  0  &   0   &  0  &   0  &   0  &   0  &   0   &  0  &   0 \\
     a_4^{1*}  &   a_4^{2*}   &  a_4^{3*}  &   a_4^{4*}  &   0  &   0  &   0  &   0  &   0  &   0  &   0  &   0  &   0  &   0   &  0  &   0 \\
     a_2^1   &  0  &   0  &   0  &   a_2^2  &   0  &   0  &   0  &   0  &   0   &  0  &   0   &  0   &  0  &   0   &  0 \\
     |a_2^1|^2 &   a_2^1a_2^{2*}  &   0  &   0  &   a_2^{1*}a_2^2  &   |a_2^2|^2  &   0  &   0  &   0  &   0   &  0  &   0  &   0  &   0  &   0  &   0 \\
     a_2^1a_3^{1*} &   a_2^1a_3^{1*}  &   a_2^1a_3^{3*}  &   0  &   a_3^{1*}a_2^2  &   a_2^2a_3^{2*}  &   a_2^2a_3^{3*}  &   0   &  0  &   0  &   0  &   0  &   0   &  0  &   0  &   0 \\
     a_2^1a_4^{1*} &   a_2^1a_4^{2*}   &  a_2^1a_4^{3*}   &  a_2^1a_4^{4*}  &   a_2^2a_4^{1*}  &   a_2^2a_4^{2*}  &   a_2^2a_4^{3*} &    a_2^2a_4^{4*}  &   0   &  0  &   0  &   0  &   0  &   0  &   0   &  0 \\
     a_3^1  &   0  &   0   &  0   &  a_3^2   &  0   &  0  &   0  &   a_3^3  &   0  &   0  &   0  &   0   &  0  &   0  &   0 \\
      a_3^1a_2^{1*} &   a_3^1a_2^{2*} &    0  &   0  &   a_3^2a_2{1*} &   a_3^2a_2^{2*}  &   0  &   0   &  a_3^3a_2^{1*} &    a_3^3a_2^{2*}&   0  &   0   &  0   &  0   &  0  &   0 \\
     |a_3^1|^2  &   a_3^1a_3^{2*}  &   a_3^1a_3^{3*}  &   0  &   a_3^2a_3^{1*}   &  |a_3^2|^2  &   a_3^2a_3^{3*}  &   0   &  a_3^3a_3^{1*}  &   a_3^3a_3^{2*}  &   |a_3^3|^2  &   0  &   0  &   0  &   0  &   0 \\
     a_3^1a_4^{1*}   &  a_3^1a_4^{2*}  &   a_3^1a_4^{3*}  &   a_3^1a_4^{4*}  &   a_3^2a_4^{1*}   &  a_3^2a_4^{2*}   &  a_3^2a_4^{3*}  &   a_3^2a_4^{4*}  &   a_3^3a_4^{1*}  &   a_3^3a_4^{2*}  &   a_3^3a_4^{3*}  &   a_3^3a_4^{4*}  &   0   &  0   &  0  &   0 \\
     a_4^1  &   0  &   0  &   0  &   a_4^2  &   0   &  0   &  0   &  a_4^3   &  0  &   0  &   0  &   a_4^4  &   0  &   0  &   0 \\
     a_4^1a_2^{1*}  &   a_4^1a_2^{2*}  &   0   &  0  &   a_4^2a_2^{1*}  &   a_4^2a_2^{2*}   &  0   &  0   &  a_4^3a_2^{1*}  &   a_4^3a_2^{2*}&   0  &   0  &   a_4^4a_2^{1*} &    a_4^4a_2^{2*}   &  0  &   0 \\
     a_3^{1*}a_4^1&   a_4^1a_3^{2*}  &   a_4^1a_3^{3*}  &   0   &  a_3^{1*}a_4^2   &  a_3^{2*}a_4^2  &   a_3^{3*}a_4^2  &   0  &   a_4^3a_3^{1*}   &  a_4^3a_3^{2*}  &   a_4^3a_3^{3*}   &  0   &  a_4^4a_3^{1*}   &  a_4^4a_3^{2*}  &   a_4^4a_3^{3*}  &   0 \\
     |a_4^1|^2 &  a_4^1a_4^{2*}  &   a_4^1a_4^{3*} &    a_4^1a_4^{4*}  &   a_4^{1*}a_4^2  &   |a_4^2|^2   &  a_4^2a_4^{3*}  &   a_4^2a_4^{3*} &  a_4^{1*}a_4^3   &  a_4^{2*}a_4^3   &  |a_4^3|^2   &  a_4^3a_4^{4*}  &   a_4^{1*}a_4^4  &   a_4^{2*}a_4^4   &  a_4^{3*}a_4^4  &   |a_4^4|^2
     \end{array}
    \right),
   $}
\end{center}

% spectra

\begin{figure}[!htp]
  \centering
  \includegraphics[width=17cm]{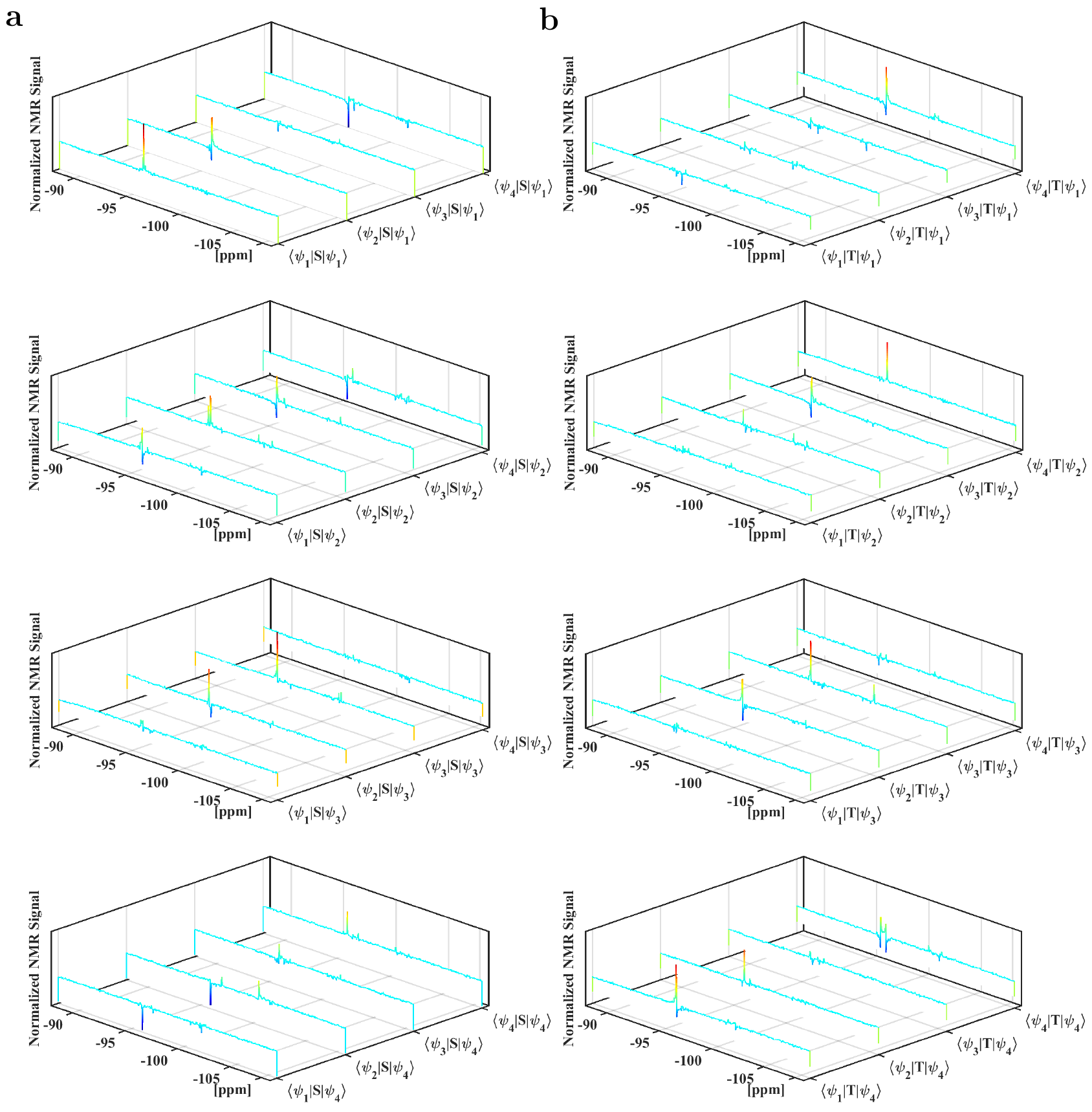}\\
  \caption{Experimental $^{19}$F spectra of measuring all elements of $S$ and $T$ matrices in the random basis $\{|\psi_i^{\text{rd}}\rangle\}$ for $h=1$.}\label{fig:8_SM}
\end{figure}
\begin{figure}[!htp]
  \centering
  \includegraphics[width=17cm]{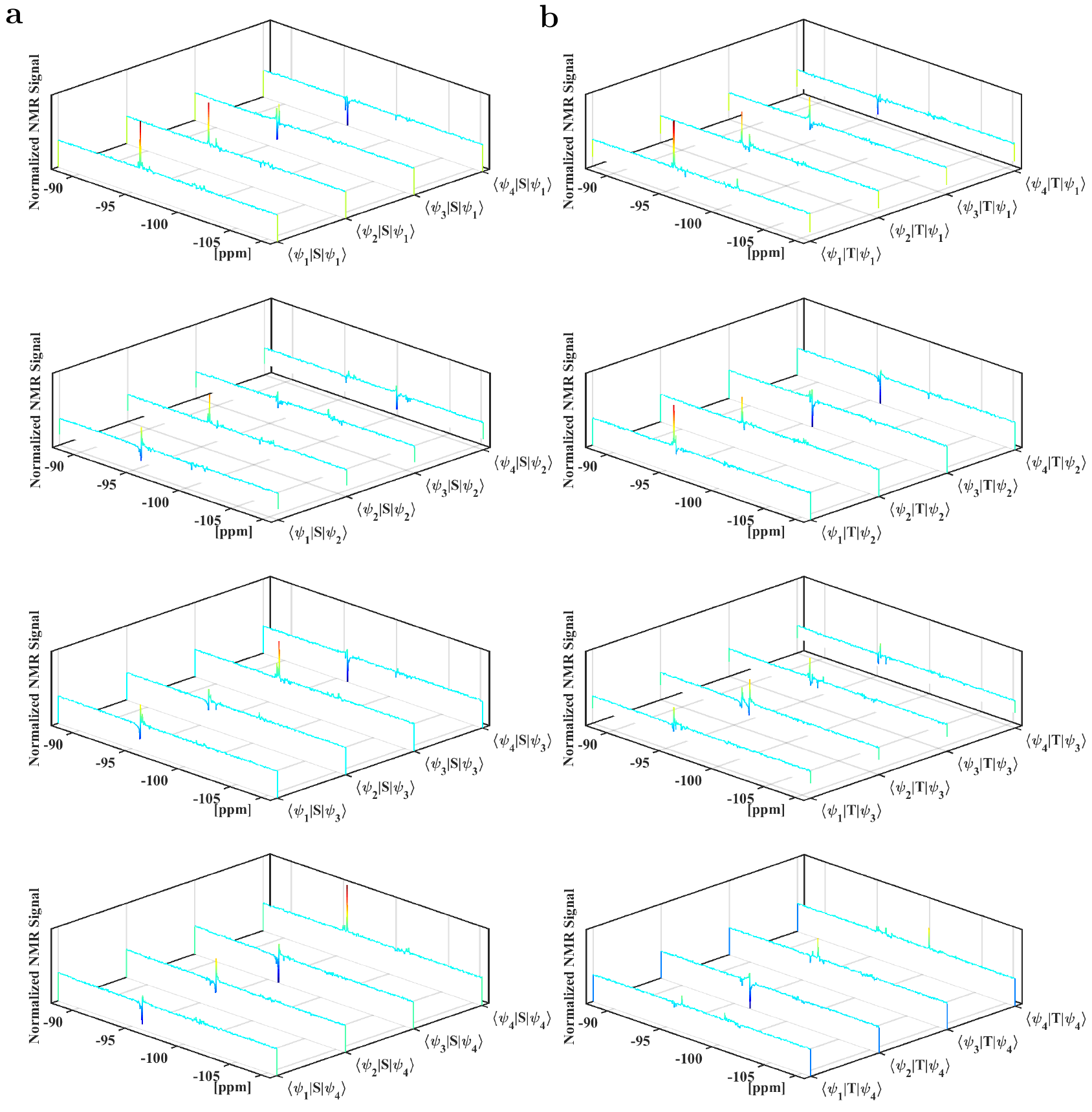}\\
  \caption{Experimental $^{19}$F spectra of measuring all elements of $S$ and $T$ matrices in the random basis $\{|\psi_i^{\text{rd}}\rangle\}$ for $h=2$.}\label{fig:9_SM}
\end{figure}
\begin{figure}[!htp]
  \centering
  \includegraphics[width=17cm]{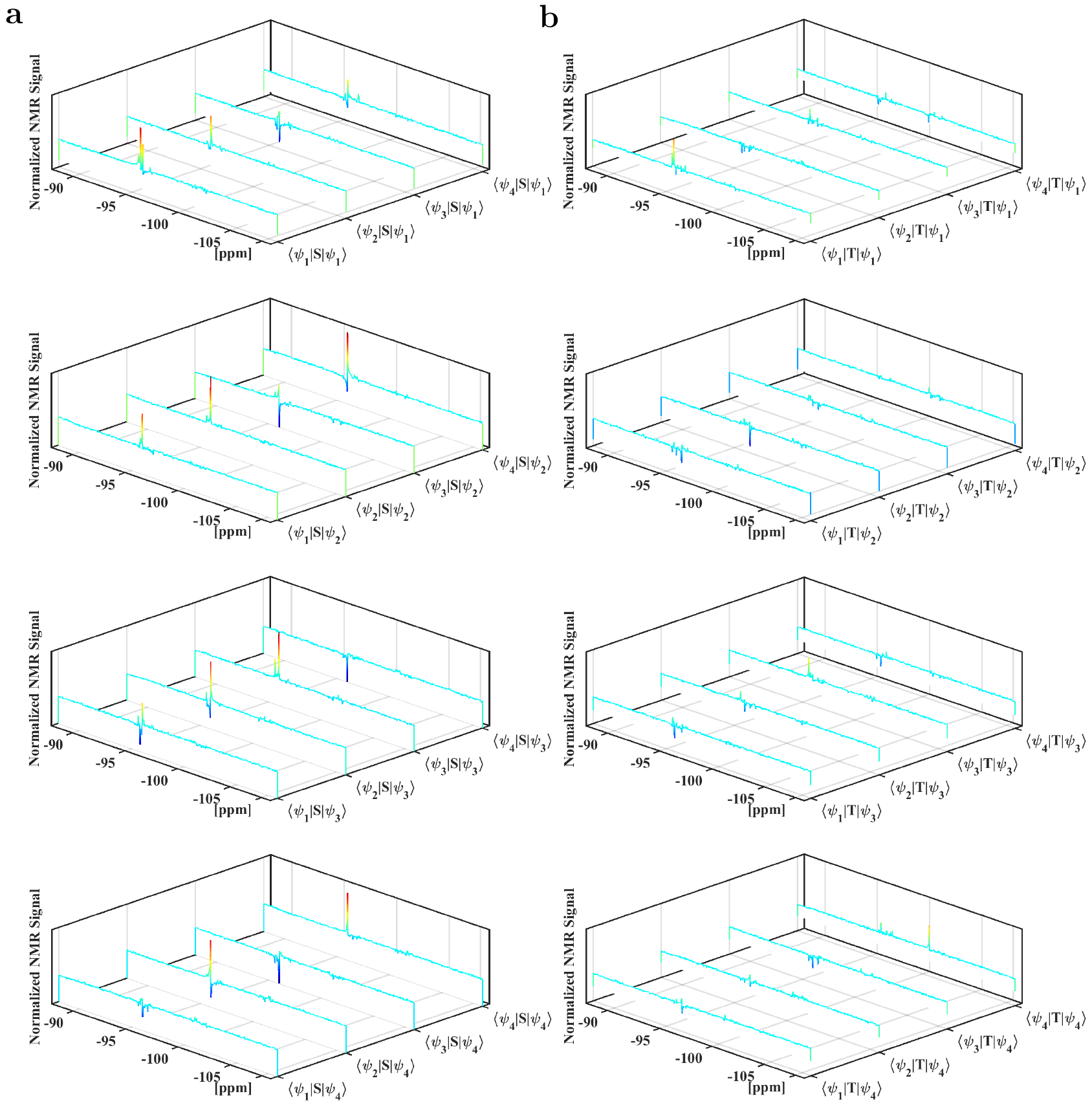}\\
  \caption{Experimental $^{19}$F spectra of measuring all elements of $S$ and $T$ matrices in the random basis $\{|\psi_i^{\text{rd}}\rangle\}$ for $h=2\sqrt{10}/3$.}\label{fig:10_SM}
\end{figure}

\end{document}